\documentclass[%
 aip,
 amsmath,amssymb,
 reprint,%
]{revtex4-2}

\usepackage{graphicx}
\usepackage{dcolumn}
\usepackage{bm}

\usepackage{bm}        
\usepackage{amsfonts}  
\usepackage{amsmath}   
\usepackage{amssymb}   
\usepackage{subfloat}  
\usepackage{float}
\usepackage{bm}

\usepackage{graphicx}
\usepackage{caption}
\usepackage{subcaption}

\usepackage[utf8]{inputenc}
\usepackage[T1]{fontenc}
\usepackage{mathptmx}
\usepackage{etoolbox}
\usepackage{multirow}

\usepackage[colorlinks=true,
linkcolor=cyan,
citecolor=cyan,
urlcolor=cyan,
]{hyperref}%

\usepackage{xcolor}

\makeatletter
\def\munderbar#1{\underline{\sbox\tw@{$#1$}\dp\tw@\z@\box\tw@}}
\makeatother

\makeatletter
\def\@email#1#2{%
 \endgroup
 \patchcmd{\titleblock@produce}
  {\frontmatter@RRAPformat}
  {\frontmatter@RRAPformat{\produce@RRAP{*#1\href{mailto:#2}{#2}}}\frontmatter@RRAPformat}
  {}{}
}%
\makeatother
\begin{document}

\preprint{AIP/123-QED}
	\title{On the unsteady aerodynamics of flapping wings under dynamic hovering kinematics}

	\author{Romain Poletti}
		\email{romain.poletti@vki.ac.be}

	\affiliation{
		von Karman Institute for Fluid Dynamics, Waterloosesteenweg 72, Sint-Genesius-Rode, Belgium
	}
        \affiliation{
        Department of Electromechanical, Systems and Metal Engineering, Ghent University,
        Sint-Pietersnieuwstraat 41, Gent, Belgium
	}
 	\author{Andre Calado}
  \altaffiliation[Now at ]{Department of Mechanical and Aerospace Engineering, The George Washington University.}
	\affiliation{
		von Karman Institute for Fluid Dynamics, Waterloosesteenweg 72, Sint-Genesius-Rode, Belgium
	}

	\author{Lilla K. Koloszar}
	\affiliation{
		von Karman Institute for Fluid Dynamics, Waterloosesteenweg 72, Sint-Genesius-Rode, Belgium
	}
    \author{Joris Degroote}
        \affiliation{
        Department of Electromechanical, Systems and Metal Engineering, Ghent University,
        Sint-Pietersnieuwstraat 41, Gent, Belgium
	}
	\author{Miguel A. Mendez}
	\affiliation{
		von Karman Institute for Fluid Dynamics, Waterloosesteenweg 72, Sint-Genesius-Rode, Belgium
	}

	\date{\today}

\begin{abstract}
Hummingbirds and insects achieve outstanding flight performance by adapting their flapping motion to the flight requirements. 
Their wing kinematics can change from smooth flapping to highly dynamic waveforms, generating unsteady aerodynamic phenomena such as leading-edge vortices (LEV), rotational circulation, wing wake capture, and added mass. 
This article uncovers the interactions of these mechanisms in the case of a rigid semi-elliptical wing undergoing aggressive kinematics in the hovering regime at $Re\sim \mathcal{O}(10^3)$. The flapping kinematics were parametrized using smoothed steps and triangular functions and the flow dynamics were simulated by combining the overset method with Large Eddy Simulations (LES).
The analysis of the results identifies an initial acceleration phase and a cruising phase. During the former, the flow is mostly irrotational and governed by the added mass effect. The added mass was shown to be responsible for a lift first peak due to the strong flapping acceleration. The dynamic pitching and the wing wake interaction generate a second lift peak due to a downwash flow and a vortex system on the proximal and distal parts of the wing's pressure side. Conversely, aerodynamic forces in the cruising phase are mainly governed by the growth and the establishment of the LEV. Finally, the leading flow structures in each phase and their impact on the aerodynamic forces were isolated using the extended Proper Orthogonal Decomposition (POD).

\end{abstract}

\maketitle






\section{Introduction} \label{sec_Intro}
 

Flapping Wing Micro Air Vehicles (FWMAVs) have shown enormous potential in demanding tasks such as rescues in confined environments, surveillance missions, pollination tasks, or even Martian surveys \cite{Phan2019,Pohly2021}. Inspired by the remarkable flight capabilities of birds and insects, honed over millions of years of evolution, the design of these MAVs seeks to replicate nature's efficiency. However, despite the significant recent advancements, modern FWMAVs are still far from being able to exploit nature's full potential \citep{Haider2021}, exemplified by the agile maneuvers of hummingbirds evading threats\cite{Cheng2016}, bats swiftly altering flight direction within two wing beats\cite{Tian2006} or owls silently surprising their prey\cite{Beratlis2020}.

The primary challenge in biomimicry lies in the need for fully characterizing and understanding the complex unsteady aerodynamics generated by flapping flight across various species \cite{Chin2016}. At the low Reynolds number regime ($Re\sim\mathcal{O}(10^2-10^4)$) typical of hovering flight, four unsteady aerodynamic mechanisms are well-known, as explained in the following paragraphs.

The leading mechanism is the flow detachment from the leading edge at high angles of attack, resulting in a Leading-Edge Vortex (LEV) on the wing's leeward side that prevents stall \cite{Ellington1984_IV, Dickinson1999, Sane2003}. The dynamics of the LEV depend on the wing's shape and kinematics\cite{Eldredge2019, Harbig2013} and much research is being conducted on the physics that drive its stability\cite{Chen2023}. Using potential flow theory, \citet{Xia2023} recently showed that the stabilization originates from the span-wise flow driven by centripetal acceleration and Coriolis forces \cite{Lentink2009}. \citet{Wabick2023} confirmed and enriched these findings using experimental data and advanced post-processing.

The second well-known phenomenon is the rotational circulation that the wing generates to preserve the Kutta condition at its trailing edge when it rapidly pitches and flaps at the same time\cite{Sane2002}. This phenomenon is called the Kramer effect\cite{Chin2016} and has drawn less attention because it is generally much weaker than the LEV. Nevertheless, \citet{Liu2020,Meng2015} have shown that rotational circulation can be responsible for large lift production in the case of fruit flies and mosquitoes. These authors estimate the force using the first moment of vorticity\cite{Wu1981,Lyu2019} applied on high-fidelity CFD data.

The third mechanism is the added mass, also known as virtual mass, generated due to the surrounding air resisting the change of momentum imparted by the wing's acceleration\cite{Sane2001,Howe1995}. The added mass generally arises at the beginning of the motion reversal as demonstrated by potential flow computations on mosquitoes\cite{Bomphrey2017}. 

The added mass is often intertwined with the last mechanism, namely the wing-wake interaction, generated due to the cyclic motion of the wing. \citet{Lee2018} classify the possible wing-wake interactions in four categories: (1) interaction with downwash/upwash flow fields, (2) formation of an impinging jet, (3) interaction with wake vortices, (4) formation of a closely attached LEV. \citet{Chakraborty2023} have further analyzed the wake using data-driven techniques such as Principal Component Analysis (PCA) to identify the region of injection of rotationality in the wake.

Other aerodynamic phenomena such as clap and fling\cite{Cheng2019,Cheng2021} or the Wagner effect\cite{Van2022} may occur during a flapping cycle. The balance between these mechanisms depends on the species, the wing geometry and kinematics. Mosquitoes\cite{Liu2020}, fruit flies\cite{Dickinson1999}, hummingbirds\cite{Song2014} hover using very different kinematics and their influence on flapping aerodynamics is not fully addressed in the literature.

Many authors have investigated the aerodynamics of flapping flight with the goal of identifying the optimal flapping laws. 
\citet{Bhat2020} classifies the literature into two broad categories: those that focus on the impact of flapping parameters while keeping a certain waveform of the flapping kinematics and those that focus on optimizing the flapping waveform while keeping prescribed bounds on macroscopic parameters such as flapping amplitude, frequency or angle of attack. Within the first category, for example, \citet{Bos2013} studied the influence of the angle of attack, the Reynolds number, and the Rossby number on the aerodynamic forces, using laminar simulations and deforming meshes on an elliptical wing and focusing on the dynamics of the LEV. Other studies can be found in the literature review from \citet{Haider2021} and \citet{Bhat2020}. Within the second category, \citet{Berman2007} analyzed the power consumption of three insect wings constrained to the hovering regime using quasi-steady models and found that the optimal performances are obtained for flapping kinematics generating a stable LEV. This work was extended by \citet{Bhat2020} using high-fidelity CFD of a fruit fly wing which showed that lift production is maximized by smooth flapping combined with rapid pitching while the efficiency is maximized by dynamic flapping and moderately smooth pitching. 

These findings motivate the interest in dynamic flapping kinematics, which is a far less explored territory. Most of the literature focuses on the dominant, quasi-steady LEV rather than on the dynamics of transient structures governing the unsteady forces at the beginning of each flapping cycle. These can have an important role, as emphasized by \citet{chen2020}, who focused on the development of the LEV formed on an impulsively started rectangular wing, and \citet{Liu2020}, who analyzed the flow dynamics in the highly dynamic case of mosquitoes. Qualitative and quantitative understanding of the flow field generated by highly dynamic kinematics are lacking and would be needed to scope the potential and the limitations of flapping flight. Moreover, the aerodynamics on dynamic flapping kinematics is mostly studied for ``low" Reynolds numbers ($Re\sim\mathcal{O}(10^2)$), that is for small insect wings. Kinematics at $Re\sim\mathcal{O}(10^3)$, such as hawkmoths or hummingbirds are comparatively less explored and yet of significant interest for the efficiency of FWMAVs, as also reported by \citet{Bayiz_2018}.

This paper aims to reveal the intertwined unsteady aerodynamic phenomena in the case of a semi-elliptical wing in highly dynamic flapping motion, considering hovering conditions at $Re\sim\mathcal{O}(10^3)$. The flow dynamics is simulated with high-fidelity CFD combining the overset method and Large Eddy Simulation (LES) and post-processed using a combination of techniques, including (1) potential flow computation to identify the contribution of added mass, (2) Q-fields and Finite-Time Lyapunov Exponent (FTLE)\cite{haller2002} computation to highlight coherent structures and (3) extended Proper Orthogonal Decomposition (POD) to quantify and visualize the impact of the unsteady mechanisms on the pressure distribution on the wing.

The rest of the article is structured as follows. Section \ref{sec_modelDef} defines the wing geometry and the investigated flapping kinematics. Section \ref{sec_method} describes the numerical settings of the simulations, including the high-fidelity solver and the potential flow solver. Section \ref{sec_Tools} describes the post-processing tools while Section \ref{sec_results} analyses the results, separating the cycle into an accelerating phase and a cruising phase. Section \ref{sec_conclusions} discusses the main outcomes of this work and suggests avenues for future works.

\section{Test case definition}\label{sec_modelDef}

\begin{figure*}[!ht]\center
     \begin{subfigure}{0.3\textwidth}
         \centering
	 \includegraphics[width=\textwidth]{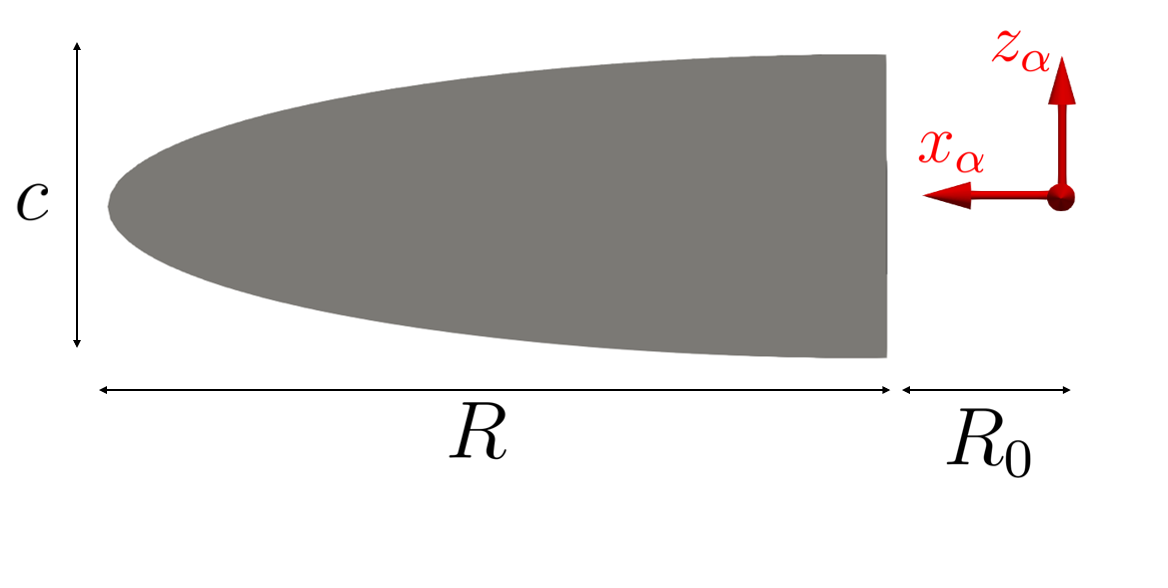}
                     \caption{ }
         \label{img_wingDiag1}
     \end{subfigure}
          \begin{subfigure}{0.5\textwidth}
         \centering
	 \includegraphics[width=\textwidth]{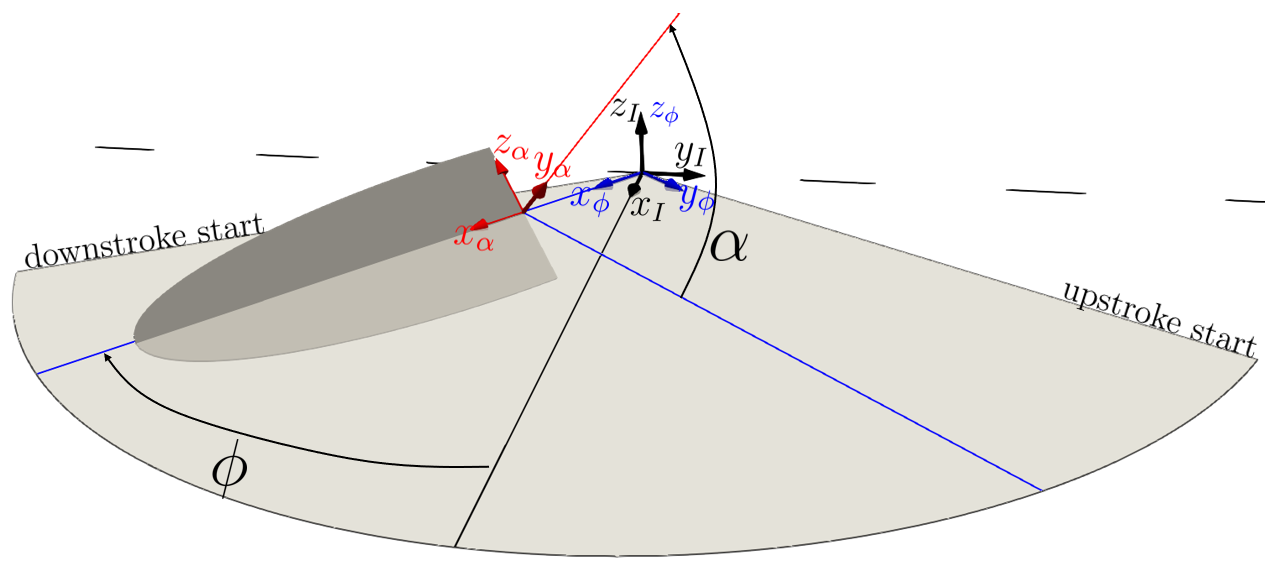}
                     \caption{ }
 \label{img_wingDiag2}    
     \end{subfigure}
     \caption{(a) Illustration of the shape of the wing and (b) its motion defined with three frames ($(xyz)_I$, $(xyz)_\phi$, $(xyz)_\alpha$) and two Euler angles (flapping angle $\phi$ and pitching angle $\alpha$).}
 \label{img_wingDiag}    
\end{figure*}


We consider a rigid semi-elliptical wing as in \citet{Lee2016}. Figure \ref{img_wingDiag} shows its characteristic dimensions and the variables governing its motion. The wing has a span $R=5$ cm, a mean chord $\bar{c}=1.5$ cm, and a uniform thickness of $b=3\%\bar{c}$. The aspect ratio $AR=R/\bar{c}=3.25$ falls within the range of hovering flyers \cite{Chin2016}. This test case serves as a canonical alternative to the countless existing platforms of hovering flyers, since the leading edge shape is the main geometrical factor influencing the wake structures\cite{Krishna2020}.
The wing is offset by $R_0=2.25$ cm from the center of rotation, along which it moves with two degrees of freedom, defined with two Euler angles and three reference frames (Figure \ref{img_wingDiag2}). The flapping angle $\phi$ is defined between the feathering axis aligned with $x_\phi$ of the stroke frame $(xyz)_\phi$ and the $x_I$ axis of the inertial frame $(xyz)_I$. The pitching angle $\alpha$ is defined between the chord normal direction $y_\alpha$ of the wing-attached frame $(xyz)_\alpha$ and $y_\phi$ of the stroke frame. The wing tip always remains in the stroke plane $(xy)_\phi$ which is always overlapping the inertial plane $(xy)_I$.
This assumption holds when studying the kinematics of hovering species because their wings show only little out-of-stroke plane motion that has negligible influence on the aerodynamic forces \cite{Ellington1984_III,Sane2002}. For forward flight, \citet{Chen2024} have recently shown the influence of the wing tip deviation from the stroke plane on the lift and the thrust.

The time variation of the flapping angle and the pitching angle follows the parametrization defined in \citet{Berman2007}:

\begin{eqnarray}
\label{eq_phi}
\phi (t) = \frac{A_{\phi}}{\arcsin (K_{\phi})}\arcsin [K_{\phi}\cos(2 \pi  ft)] \\
\label{eq_alpha}
\alpha (t) = \frac{A_{\alpha}}{\tanh (K_{\alpha})}\tanh [K_{\alpha}\sin(2 \pi  ft)] \
\end{eqnarray}

The frequency is set to $f=20$ Hz, the flapping amplitude to $A_\phi=60^{\circ}$ and the pitching amplitude to $A_\alpha=45^{\circ}$ as inspired by typical hovering flights such as crane flies, honey bees, hawkmoths, and hummingbirds\cite{Ellington1984_III,Liu2009,Song2014}. The shape factors $K_{\phi}$ and $K_{\alpha}$ define the kinematics waveform. These factors adapt the motion profiles from sinusoidal ($K$'s low) to triangular and square waveforms for the flapping and pitching motion respectively ($K$'s high). 

This work considers a highly dynamic case with $(K_\phi,K_\alpha)=(0.99,10)$ as the reference case and two cases with $(K_\phi,K_\alpha)=(0.5,10)$ and $(K_\phi,K_\alpha)=(0.99,5)$ to compare the effects of the extreme dynamics. Figure \ref{img_motionParametrization} shows the time evolution ($t'=tf)$ of $\phi(t)$ and $\alpha(t)$ for the three investigated flapping kinematics, along with the associated angular velocities and accelerations during one flapping period. 
The plots are shaded in gray in the upstroke portion, to distinguish it from the downstroke one. In the most dynamic case, the wing starts at rest, with its chord aligned with $z_I$, and rapidly accelerates to a constant flapping velocity and to the prescribed pitching amplitude. The wing keeps these values towards the stroke end, where it decelerates and flips (stroke reversal). The wing starts then the next stroke that is identical to the first one but with opposite flapping rotations.  

The proposed kinematics result in highly unsteady aerodynamic phenomena that are much less explored in the literature than quasi-steady and smooth sinusoidal kinematics. Based on this kinematics, the chord-based Reynolds number is defined as $Re = U_{ref} \bar{c}/\nu\approx4000$ for which the reference velocity is $U_{ref}=4fA_{\phi}R_2$ at the radius of second moment of area $R_2=\sqrt{\int_{R_0}^{R_0 + R}c(r)r^2dr/S}$ \cite{Lee2016}. This $Re$ value belongs to the largest hovering animals such as hummingbirds or hawkmoths. 
The Rossby number is $Ro = R_2/\bar{c}=2.97$, which is the lower limit of the typical range of hovering flyers $Ro\sim[3,4]$ \cite{Chin2016}. The higher this number, the closer the flapping rotation approaches a pure translating motion which decreases the LEV stability\cite{Lee2016}.

\section{Numerical tools}\label{sec_method}

Section \ref{sec_CFD} reports on the numerical setup of the LES and overset simulations on which the aerodynamic analysis in this work is based. Section \ref{sec_potentialFlow} reports on the potential flow computations that were used to identify viscous and non-circulatory effects. 

\subsection{LES simulation with the overset method} \label{sec_CFD}
The simulations were carried out in the open-source software OpenFOAM, in particular with the solver called \textit{overPimpleDyMFoam}. 

This is a finite volume code used for the incompressible Navier-Stokes equations using the PIMPLE algorithm for the pressure-velocity coupling.

\begin{figure}[htb] \center
 \includegraphics[width=0.5\textwidth]{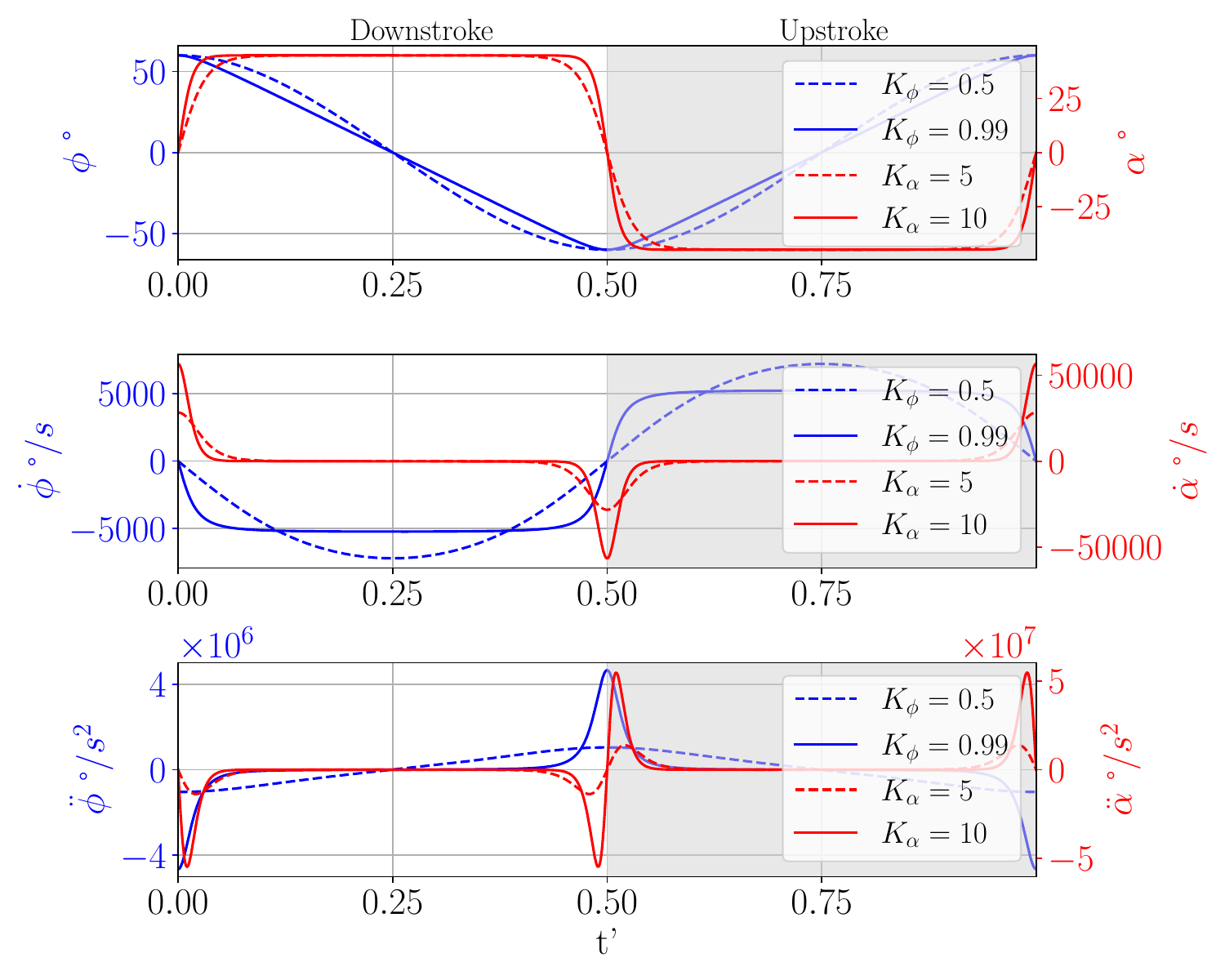}
 \caption{Flapping and pitching angles (top), angular velocities (middle) and angular accelerations (bottom) according to the parametrization in equation \eqref{eq_phi} and \eqref{eq_alpha} for the three test cases investigated in this work, i.e. $(K_\phi,K_\alpha)=(0.99,10)$, $(K_\phi,K_\alpha)=(0.5,10)$ and $(K_\phi,K_\alpha)=(0.99,5)$.}
 \label{img_motionParametrization}
\end{figure}

\begin{figure}[htb] \center
 \includegraphics[width=0.45\textwidth]{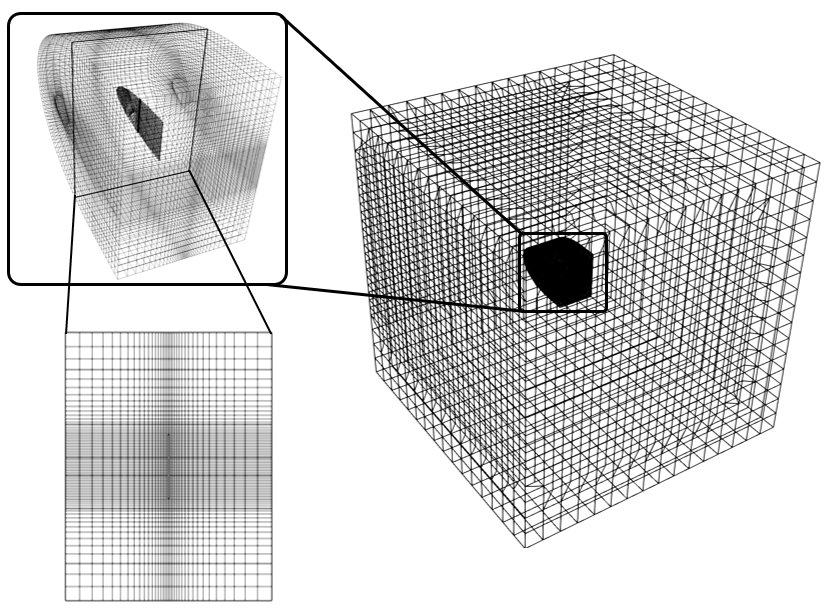}
 \caption{The background grid, the component grid (zoom), and a spanwise slice of the latter.}
 \label{img_CFD}
\end{figure}

The computational grid was adapted to the wing displacement using the overset method. This method, also known as the chimera technique \cite{Steger1983}, is used together with the Arbitrary-Lagrange-Euler (ALE) formulation, whereby the convective velocity of the momentum equation is modified to account for the displacement of the grid\cite{Hirt1974}:

 \begin{eqnarray} \label{eq_NS}
\frac{\partial \bm{u}}{\partial t} + (\bm{u} - \bm{u}_g) \cdot \bm{\nabla} \bm{u}   = -  \bm{\nabla}p +  \nu \bm{\Delta} \bm{u}\,,
\end{eqnarray} where $\bm{u}=[u_x,u_y,u_z]$ is the flow velocity, $p$ is the kinematic pressure, $\nu$ the kinematic viscosity, and $\bm{{u}_g}$ is the grid velocity. The grid is composed of a component grid that moves with the wing and a background grid that is fixed ($\bm{{u}_g}=\bm{0}$). These grids overlap and are shown in Figure \ref{img_CFD}.
The flow equations are solved simultaneously on both grids (strong coupling \cite{Hadzic2005}), exchanging pressure and velocity fields through interpolation\cite{Pynaert2022} using the inverse distance scheme.
More details about the numerical solver are provided in \citet{Poletti2023}.

Figure \ref{img_CFD} shows the designed grid consisting of $430k$ hexahedral cells in total. The component grid has a C-grid topology with $n_{cells} = 130k$, with the first cell having a height of 3$\%$ of the chord. The cell height grows with an expansion ratio of $1.1$ until the component grid boundary. The boundaries of the component grid are one chord length away from the wing, following literature standards in flapping wing simulations \cite{Badrya2017,Nakata2015}. A smaller grid would reduce the computational time but would be prone to larger numerical errors because the interpolation region would move closer to the wing, where stronger gradients are expected.  
The background grid is a 20-chord cube which prevents boundary influence on the wing \cite{Badrya2017}. 
A refinement zone is also set along the wing trajectory such that the cell size of the background and component grid coincide at their interface, as this minimizes inter-grid interpolation errors \cite{Hadzic2005}. \citet{Calado2023} used the same grid and showed that the component grid resolution has the largest influence due to the strongest flow gradients. Reducing the component grid density from $430k$ to $130k$ cells induces a deviation of less than 1$\%$ in the mean lift, hence proving the grid independence of the selected grid. The cell count of the background grid was also reduced from 950k cells to 300k cells to preserve similar cell sizes at the interface between the two grids. The complete results of the grid study are found in \citet{Calado2023}.

Concerning the numerical schemes, second-order central Gauss schemes were used to discretize the spatial terms with limiters in the direction of the most severe gradients. 
A backward, second-order, implicit scheme was used for the time marching and the simulations were performed with adaptative time steps capped by a maximum Courant number of 1.
Neumann conditions were applied to the velocity field, and Dirichlet conditions were applied to the pressure on the background grid boundaries to model the hovering regime for which the fluid is initially at rest.

\begin{figure}[htb] \center
 \includegraphics[width=0.45\textwidth]{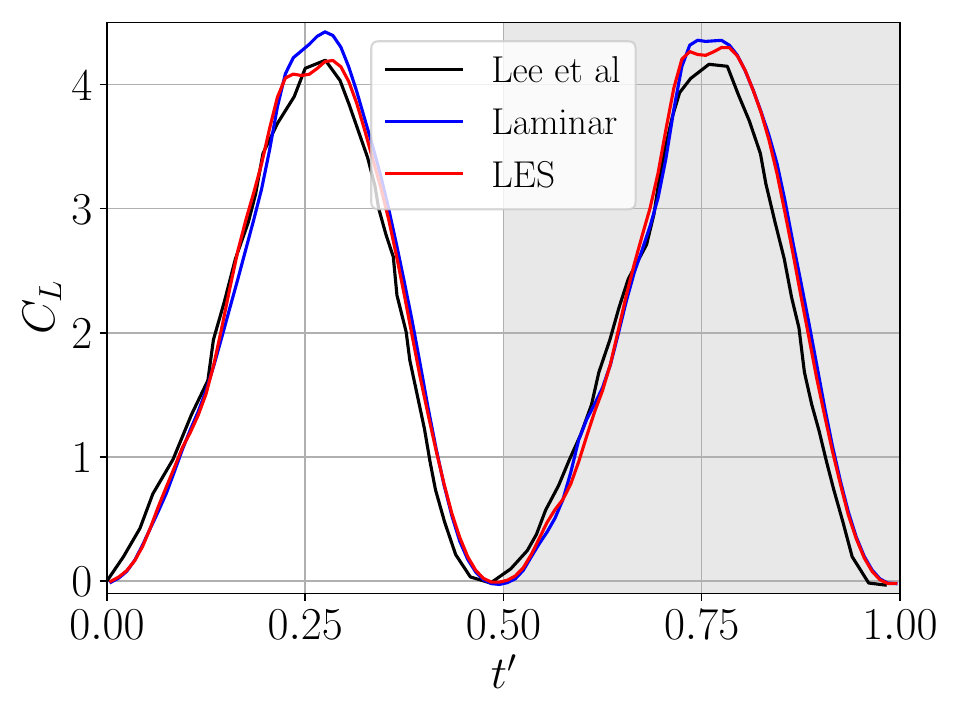}
 \caption{Evolution of the lift coefficient during a flapping with $Re\approx 4000$, $A_\phi = 60^{\circ}$, $A_\alpha = 45^{\circ}$, and $(K_\phi,K_\alpha) = (0.01,0.01)$ for a simulation using a laminar solver (blue line) a LES solver (red line). The black line shows the prediction in \citet{Lee2016}.}
 \label{img_liftValidation}
\end{figure}

Since the Reynolds number ($Re \sim \mathcal{O}(10^3)$) in these simulations lies on the theoretical limit of the transition regime, LES with the dynamic turbulent kinetic sub-grid model was used. LES aims to model the scales that cannot be resolved on the potentially too coarse grid, unlike similar literature works that perform simulations without turbulence model\cite{Bos2013,Cai2021,Badrya2017}.
Nevertheless, the influence of turbulence modeling was found to be minor, as revealed in Figure \ref{img_liftValidation}. This figure plots the lift coefficient during one flapping cycle in dimensionless form, comparing the laminar and LES simulations with the results from the laminar simulations by \citet{Lee2016}, who used a sliding mesh approach. 
The pressure coefficient and lift coefficient (along $z_\phi$) are computed as:



\noindent\begin{minipage}{.5\linewidth}
\begin{equation}\label{eq_Cp}
C_p = \frac{\rho (p-p_\infty)}{0.5\rho U_{ref}^2}
\end{equation}
\end{minipage}%
\begin{minipage}{.5\linewidth}
\begin{equation}\label{eq_CL}
C_L = \frac{\int_S (\rho p + \bm{\tau}) z_\phi dS}{0.5\rho SU_{ref}^2},
\end{equation}
\end{minipage}

\vspace{0.2cm}
\noindent where $\rho$ is the air density, $p_\infty$ is the static pressure in the far field, $\bm{\tau}$ is the shear stress tensor, and $S$ is the wing surface. A root-mean square error below 2$\%$ was observed, thus validating the numerical approach.

\vspace{-0.8cm}
\subsection{Potential flow simulations} \label{sec_potentialFlow}

Potential flow computations are used to isolate the added mass influence from vorticity and skin friction contributions. \citet{Howe1995} and \citet{Menon2021} proved that these three contributions can be formulated into uncorrelated and physically interpretable mechanisms that add up to the total force. Similarly to \citet{Liu2020} and \citet{Menon2021}, the potential flow is solved such that $\bm{u_p}=\nabla \Phi$, with $\Phi$ the potential function satisfying the Laplace equation: 

\begin{equation} \label{eq_PHI}
\nabla^2 \Phi = 0,
\end{equation}

with the boundary conditions:

\begin{equation} \label{eq_BCPHI}
\left\{ 
\begin{array}{ll}
\bm{\nabla} \Phi = \bm{0} & \text{outer boundaries} \\
\frac{\partial \Phi}{\partial n} = \bm{u}_w(\bm{x}_w,t) \cdot \bm{n}_w(\bm{x}_w,t) & \text{wing surface}
\end{array}
\right.
\end{equation} 
The first condition imposes zero velocity at the external boundaries of the domain. The second condition enforces the non-permeability of the wing surface, with $\bm{u}_w(\bm{x}_w,t)$ the local wing velocity at the wing surface, identified by the set of points $\bm{x}_w$ and $\bm{n}_w$ the local normal to the wing surface. 

Since equation \eqref{eq_PHI} is linear, the pitching and flapping boundary conditions are imposed in distinct simulations to $\bm{u}_w$, giving two potential fields that are added up to reconstruct the total flow field. This approach allows to reveal unequivocally the pattern attributed to each motion.  

The velocity potential $\Phi$ was computed in a quasi-steady fashion: equation \eqref{eq_PHI} was solved independently for each time step based on the instantaneous wing velocity $\bm{u_w}$. The overset method is not needed and the potential flow is solved in the wing-attached frame. The computations were carried out in OpenFOAM, hence using the finite volume method, with a second-order scheme for the Laplacian operator. 

Based on the time series of the velocity potential, the unsteady Bernoulli equation gives the lift force as: 

\begin{equation} \label{eq_fPot}
C_{L,p} = \frac{\int_S [p_\infty - \frac{\partial \Phi}{\partial t} - \frac{1}{2}||\bm{\nabla} \Phi||^2] z_\phi dS}{0.5\rho SU_{ref}^2}\,,
\end{equation} where the integration of the last term is zero due to the small wing thickness.




\section{Post-processing tools}\label{sec_Tools}

\subsection{Coherent structure identification}  \label{sec_FTLE}

Both Eulerian and Lagrangian coherent structures were analyzed in the flow field produced by the investigated flapping wing. 
The Eulerian structures were identified using the Q-criterion, as it is common in the literature of flapping wing aerodynamics\cite{Bos2013, Harbig2013,Bhat2020}. The Q field is the second invariant of the velocity gradient tensor:

\begin{equation} \label{eq_Q}
Q = \frac{1}{2}[||\bm{\Omega}||^2 - ||\bm{S}||^2],
\end{equation}
where $\bm{\Omega}=0.5[\bm{\nabla \bm{u}} -  \bm{\nabla}\bm{u}^T]$ is the spin tensor and $\bm{S}=0.5[\bm{\nabla}\bm{u} +  \bm{\nabla}\bm{u}^T]$ is the strain rate tensor. Although the Q fields can effectively identify coherent patterns, this quantity is not frame-invariant\cite{haller2015}. Therefore, different vortical structures are identified in the wing frame or in the inertial frame.

This ambiguity is here resolved using Lagrangian coherent structures (LCS). These are identified by following the patterns produced by ideal tracer particles, and particularly by identifying the lines where these tend to separate or gather, forming transport barriers of the flow. 
The LCS are defined as the ridges of the Finite-Time Lyapunov Exponents fields (FTLE) and include information on the history of the flow. To briefly introduce the key steps for identifying the LCS, let $\bm{x}_i=\bm{x}(t_i)=[x_i,y_i,z_i]$ denote the set of coordinates of a single particle in an infinite set injected in the flow at time $t=t_i$. "Ideal" here means that the particles follow the flow with no lag. The flow map $\bm{F}_{t_i}^{t_f}(\bm{x}_i)$ acting on the initial positions allows for tracking the particles up to $t_f=t_i+T_o$, with $T_o$ the observation time, to retrieve the final position $\bm{x}_f=\bm{x}(t_f)=[x_f,y_f,z_f]$:

\begin{equation} \label{eq_flowMap}
\bm{x}_f=\bm{F}_{t_i}^{t_f}(\bm{x}_i) := \bm{x}_i + \int_{t_i}^{t_f}\bm{u}(\bm{x}(\tau),\tau)d\tau\,.
\end{equation}

Defining the Jacobian of the flow map as 

\begin{equation}
\bm{J}_{t_i,t_f}=\frac{d \bm{F}_{t_i}^{t_f}}{d \bm{x}_i}= \begin{bmatrix}
\partial_{x_{i}} x_f & \partial_{x_{i}} y_f  & \partial_{x_{i}} z_f\\
\partial_{y_{i}} x_f & \partial_{y_{i}} y_f  & \partial_{y_{i}} z_f\\
\partial_{z_{i}} x_f & \partial_{z_{i}} y_f  & \partial_{z_{i}} z_f
\end{bmatrix}\,,
\end{equation}

and the Cauchy-Green deformation tensor as 

\begin{equation}\label{eq_G}
\bm{\Delta}_{t_i,t_f} = \bm{J}^T_{t_i,t_f} \bm{J}_{t_i,t_f}\,,
\end{equation}

the FTLE field is given by :

\begin{equation}\label{eq_FTLE}
\bm{\sigma}_{t_i,t_f} = \frac{1}{T_o} \mbox{log}\sqrt{\lambda_{max}(\bm{\Delta}_{t_i,t_f})},
\end{equation} where $\lambda_{max}(\bm{\Delta}_{t_i,t_f})$ is the largest eigenvalue of $\bm{\Delta}_{t_i,t_f}$.
The FTLE field is 3D and contains the scaled maximum eigenvalue of the deformation tensor for each particle. A large $\bm{\sigma}$ indicates regions where trajectories are prone to stretch \cite{krishna2019}. If the time integration bounds in equation \eqref{eq_flowMap} are reversed, the largest (negative) FTLE shows where the particles separate the most backward in time (i.e. showing the coherent structure in forward time). 

Computing the time evolution of the FTLE during a flapping cycle is extremely cumbersome. The computation of the time evolution FTLE requires computing the flow map \eqref{eq_flowMap} and the field \eqref{eq_FTLE} for each time $t_i$ in the flapping period.
The time series is then played forward in time to highlight the attracting lines, which can be seen as boundaries of the coherent structures where part of the flow is trapped. The present work identifies it via contour plots (see \citet{haller2013} for analytic techniques).

\begin{figure}[htb] \center
 \includegraphics[width=0.45\textwidth]{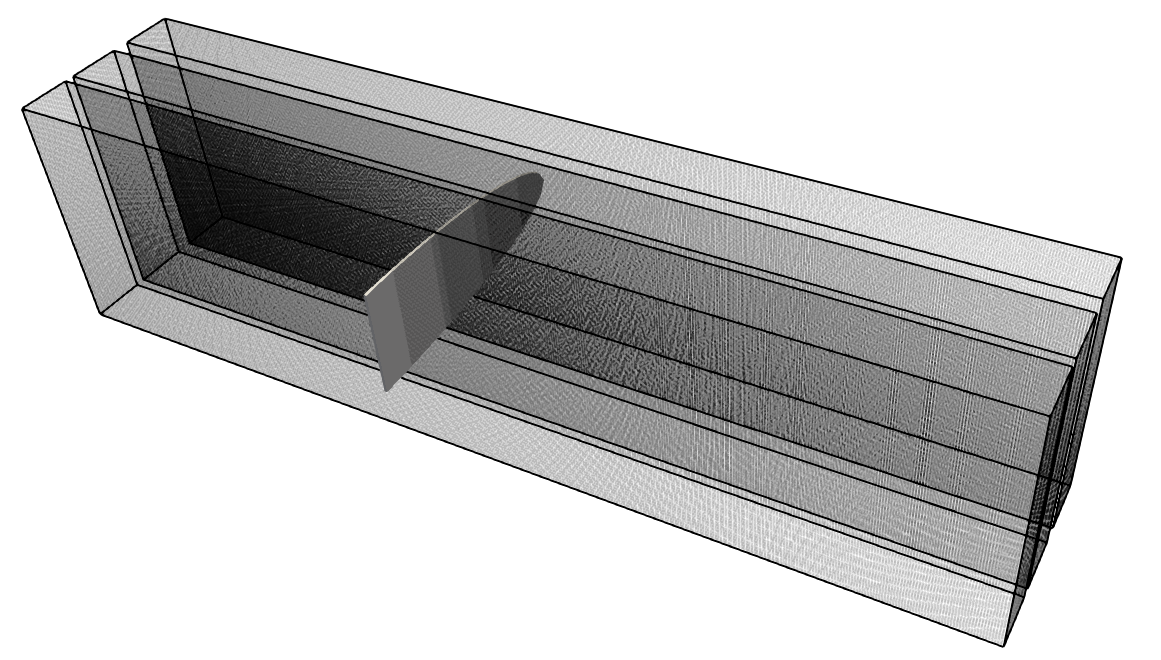}
 \caption{View of the three volumes where particles are initialized and tracked to compute the FTLE fields.}
 \label{p_initial}
\end{figure}

The computation collects a set of $n_I$ velocity fields (snapshots) which are used to integrate the particle trajectories in equation  \eqref{eq_flowMap}. For each instant $t_i$, a total of $n_p=4.5\cdot 10^6$ particles were distributed equally within three volumes along the wing span, centered at locations $R_2$, $R_2 + c$, and $R_2 - c$ as shown in Figure \ref{p_initial}. The time discretization uses a trapezoidal scheme with $d\tau=1/(400f)$ and $T_o=1/(8f)$, while an 8-point linear scheme is used for the spatial interpolation of the velocity fields. The parameters for these computations were tuned seeking a compromise between computational cost and accurate evaluation of lasting structures in the wake\cite{Krishna2018,krishna2019}. At the beginning of each FTLE calculation, the volume defining the grids of the initial particles is re-initialized and placed to have one of the edges parallel to the $y_\phi$ axis, using the rotation matrix:

\begin{equation}
    R(\phi) = \begin{bmatrix}
\cos(\phi) & -\sin(\phi)  & 0\\
\sin(\phi) & \cos(\phi) & 0 \\
0 & 0 & 1
\end{bmatrix}
\end{equation}

This allows to significantly reduce the computation time with minimal loss of details on the vortical structures.

\subsection{Extended Proper Orthogonal Decomposition (EPOD)}

By Extended POD (EPOD)\cite{Boree2003}, one usually refers to the practice of projecting (correlating) a dataset $\bm{D}_1$ onto (with) the POD modes of another dataset $\bm{D}_2$.
This allows to reveal the patterns in $\bm{D}_2$ that are most correlated with $\bm{D}_1$. In this work, we use this framework to identify the leading structure in the velocity field of flow around the flapping wing and correlate these with the pressure fields on the wing's surface and thus the associated aerodynamic loads.

The POD of the velocity field was computed using the software package MODULO\cite{Ninni2020,Modulo_2}. A total of $n_t=2048$ snapshots of the relative velocity fields $\bm{U}_r=\bm{U}-\bm{U}_g$ were saved with a uniform time stepping on the component grid during $8$ flapping cycles. 
Only the component grid is used so that the domain contains only fluid cells (see \citet{Poletti2023}) and the velocity is transformed into the wing frame to conform with the POD functioning recalled hereafter. 

The POD computation requires restructuring the datasets into a single snapshot matrix. For each time step, the flow field is reshaped into a single column vector $\bm{d}_{v,i} \in \mathbb{R}^{n_s \times 1 }$ which stacks vertically the three velocity components ($n_s = 3n_{cells}$). The snapshot matrix collecting all these vectors in time-ascending order is denoted as $\bm{D}_v \in \mathbb{R}^{n_s \times n_t}$:

\begin{equation}
    \bm{D}_v = \big[\bm{d}_{v,1},...,\bm{d}_{v,i},...,\bm{d}_{v,n_t}\big]\,.
\end{equation}

In these settings, the POD can be written as a matrix factorization of the form 

\begin{equation} \label{eq_svd}
    \bm{\tilde{D}}_{v,n_r} = \bm{\Phi}_{v} \bm{\Sigma}_{v} \bm{\Psi}_{v}^T =\sum^{n_r}_{r=1} \sigma_{v,r}\phi_{v,r}\psi^T_{v,r} 
\end{equation} where $\tilde{D}_{v,n_r}$ denotes the rank $n_r$ approximation of the matrix $\bm{D}_v$ using the leading $n_r$ POD modes. These have spatial structures $\bm{\Phi}_v=[\phi_{v,1},\phi_{v,2}\dots] \in \mathbb{R}^{n_s \times n_r}$, temporal structures $\bm{\Psi}_v=[\psi_{v,1},\psi_{v,2},\dots]\in \mathbb{R}^{n_t \times n_r}$ and amplitudes $\bm{\Sigma}=\mbox{diag}(\sigma_{v,1},\sigma_{v,2}\dots)\in\mathbb{R}^{n_r\times n_r}$. The POD minimizes the norm of the $l_2$ error of the approximation $\bm{D}-\tilde{\bm{D}}$. The reader is referred to \citet{Mendez2023} for more details on the POD computations. Using the methods of snapshots\cite{Sirovich1987}, the computation of the matrices in equation \eqref{eq_svd} is carried out by first computing the temporal structures $\bm{\Psi}$, which are eigenvectors of the temporal correlation matrix $\bm{K}[i,j] = \big<\bm{d}_i,\bm{d}_j\big>_{S}$ where $\big<\cdot\big>_{S}$ denotes the weighted inner product in space taking into consideration that the samples are available on a non-uniform grid. The amplitudes $\bm{\Sigma}$ in equation \eqref{eq_svd} are the square root of the eigenvalues of $\bm{K}$ and hence the spatial structures are computed as 

\begin{equation}
\label{Proj_POD}
\bm{\Phi}_v=\bm{D}_v\bm{\Psi}_v\bm{\Sigma}_v^{-1}\,. 
\end{equation}

Equation \eqref{Proj_POD} is a projection of the dataset $\bm{D}_v$ on the temporal basis $\bm{\Psi}_v$. To identify the spatial structures in the pressure field on the wing's surface, the same projection \eqref{Proj_POD} can be carried out on the snapshot matrix $\bm{D}_p=[\bm{d}_{p,1},.. \bm{d}_{p,n_t}]\in\mathbb{R}^{n_{ss}\times n_t}$ collecting the snapshot vector $\bm{d}_{p,i}\in\mathbb{R}^{n_{ss}}$ obtained by reshaping the pressure field $p(\bm{x}_w,t_i)$ sampled on the wing surface $\bm{x}_w$ containing $n_{ss}$ cells at time step $i$: 

\begin{equation}
\label{Proj_POD_P}
\bm{\Phi}_p=\bm{D}_p\bm{\Psi}_v\bm{\Sigma}_v^{-1}\,. 
\end{equation}

The spatial structures identified in the columns of $\bm{\Phi}_v$ in equation \eqref{Proj_POD} and the columns of $\bm{\Phi}_p$ in equation \eqref{Proj_POD_P} are then to be reshaped back to their original shape for plotting purposes.



\begin{figure}[!htb]\center
	 \includegraphics[width=0.45\textwidth]{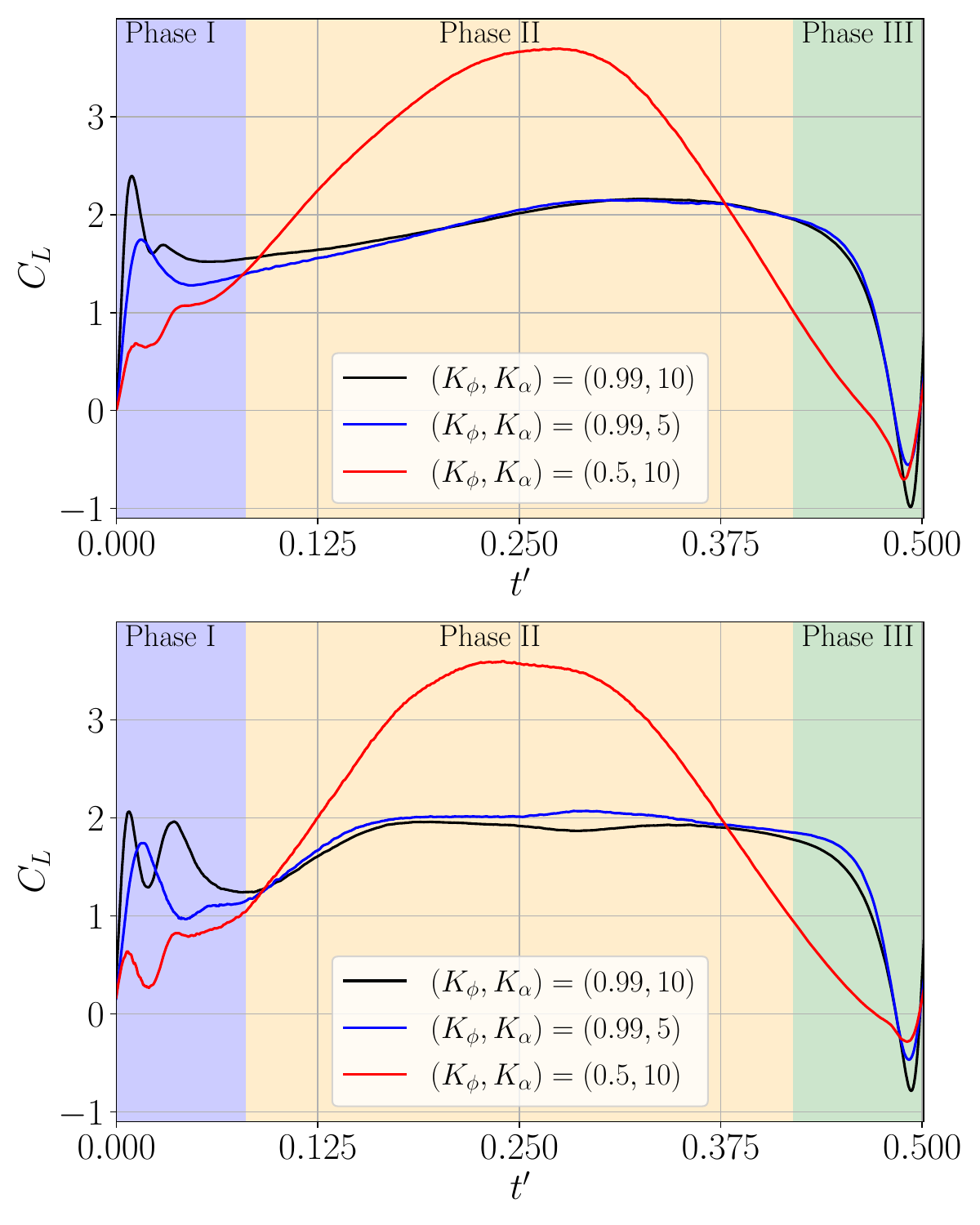}
     \caption{Influence of the wing kinematics $(K_\phi,K_\alpha)$ on the lift during the 1st downstroke (top) and 5th downstroke (bottom).}
 \label{img_liftKphiVar}    
\end{figure}

\begin{figure*}[!ht]\center
	 \includegraphics[width=0.96\textwidth]{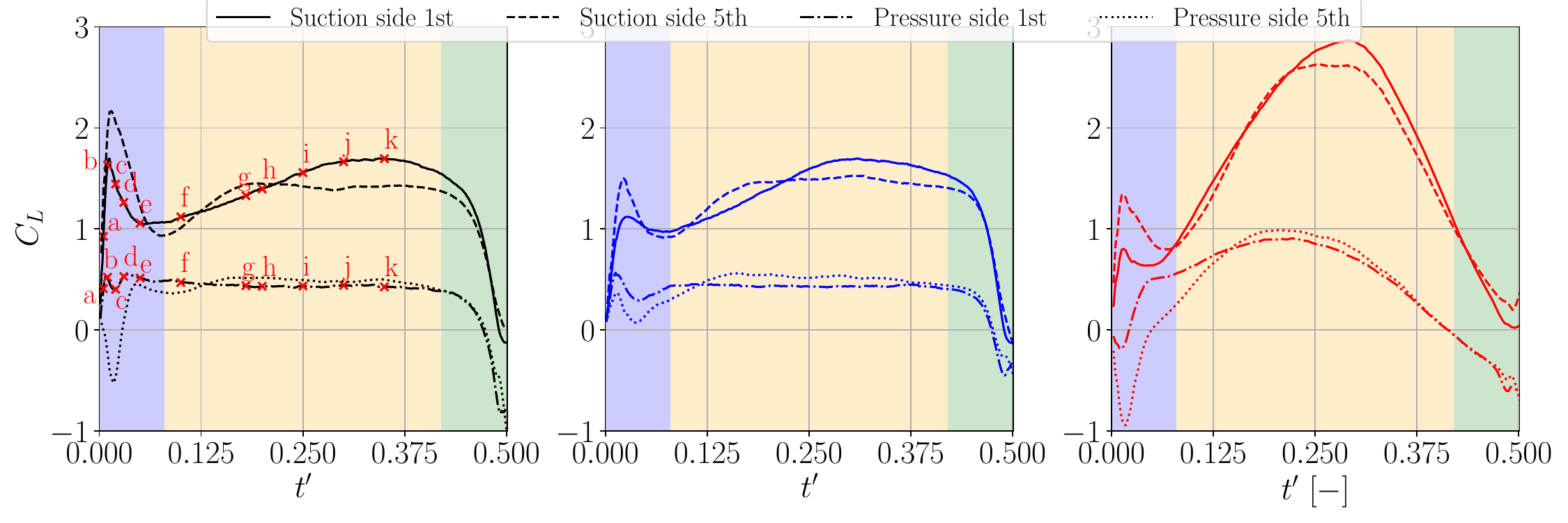}
     \caption{Lift on the suction side and pressure side of the wing during the 1st and 5th downstroke for the three test cases: $(K_\phi,K_\alpha)=(0.99,10)$ (left, in black), $(K_\phi,K_\alpha)=(0.99,5)$ (center, in blue) and $(K_\phi,K_\alpha)=(0.5,10)$ (right, in red). Red markers correspond to instants that are analyzed in separate figures: (a) t'= 0.005, (b) t'= 0.01, (c) t'= 0.02, (d) t'= 0.03, (e) t'= 0.05, (f)  t'= 0.1, (g) t'= 0.18, (h) t'= 0.2, (i)  t'= 0.25,
       (j)  t'= 0.3  ,(k) t'= 0.35.}
 \label{img_liftSide}    
\end{figure*}

\section{Results}\label{sec_results}


Figure \ref{img_liftKphiVar} shows the lift coefficient during the downstroke phase of the first (top) and fifth cycle (bottom) for the three investigated flapping kinematics. The first cycle is characterized by a distinctive transient flow as the wing moves in quiescent air, while the fifth one is characteristic of a fully developed flapping cycle, with the wing interacting with the flow structures produced by the previous cycles.   
Several authors have shown that the full establishment of the flow is achieved by the fifth cycle\cite{Badrya2017,Bos2013,Cai2021,Sun2002}.

Focusing on the most dynamic case with $(K_\phi, K_\alpha)=(0.99,10)$, the stroke is divided into an acceleration phase (phase I: $0<t'\le t'_a$) and a cruising phase (phase II: $t'_a<t'\le 0.5-t'_a $) for which $t'_a=0.08$ satisfies $\ddot{\alpha}=0.05 \% \text{max}(|\ddot{\alpha}|)$. Phase III ($t'> 0.5-t'_a $) is not analyzed in this article and this partitioning was inspired by \citet{Krishna2018, Liu2020}. The flow dynamics in each of these phases is analyzed in a dedicated subsection in the following, and compared to the smoother kinematics with $(K_\phi, K_\alpha)=(0.99,5)$ and $(K_\phi, K_\alpha)=(0.5,10)$.

The discussion in the following subsections is also built around the results in Figure \ref{img_liftSide}, which shows the lift coefficient on both sides of the wing for the two selected cycles. These were computed from equation \eqref{eq_CL}, considering the pressure and shear stress only on the bottom surface ("pressure side") and only on the upper surface surface ("suction side"). A moving average has also been applied to all the lift curves presented in this work to smooth out the oscillations resulting from the overset interpolations.

Figure \ref{img_liftSide} also shows red markers from (a) $t'=0.005$ to (k) $t'=0.035$. All of the following figures presenting a flow field are always shown at one of these times so that the reader can easily relate lift trends to flow field events.


\begin{figure*}[!ht]\center
\begin{subfigure}{\textwidth}
    \begin{minipage}{0.05\textwidth} 
        \captionsetup{justification=raggedright,singlelinecheck=false,format=hang}
        \caption{}
         \includegraphics[width=\textwidth]{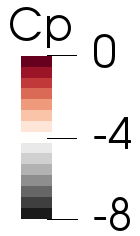}
        \label{img_psuction_1st}
    \end{minipage}%
    \begin{minipage}{0.95\textwidth} 
        \centering
        \includegraphics[width=\textwidth]{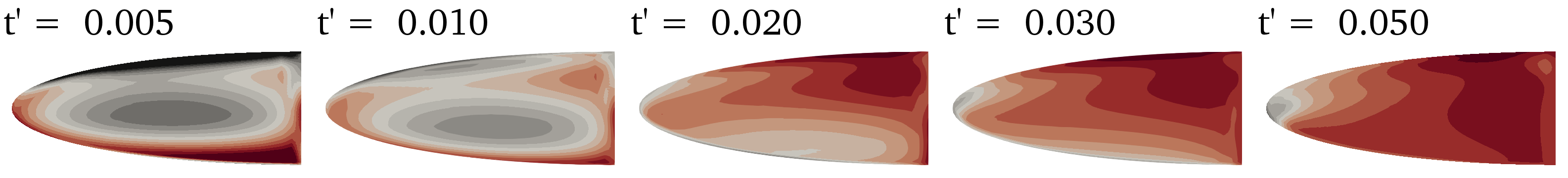}
    \end{minipage}
\end{subfigure} 
\begin{subfigure}{\textwidth}
    \begin{minipage}{0.05\textwidth} 
        \vspace{-0.3cm}
        \captionsetup{justification=raggedright,singlelinecheck=false,format=hang}
        \caption{}
         \includegraphics[width=\textwidth]{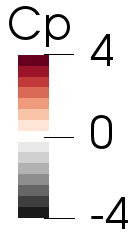}
        \label{img_ppressure_1st}
    \end{minipage}%
    \begin{minipage}{0.95\textwidth} 
        \centering
        \includegraphics[width=\textwidth]{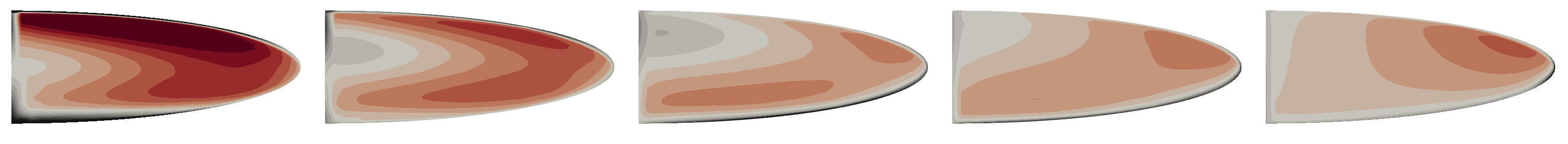}
    \end{minipage}
\end{subfigure} 
        \centering
\caption{Pressure on (a) the suction side and (b) pressure side of phase I during the 1st cycle of a viscous simulation at instants corresponding to the first five markers of Figure \ref{img_liftSide}.}
\label{img_addM_pressureField}
\end{figure*}

\subsection{Phase I: Acceleration}\label{sec_phaseI} 

At the beginning of the stroke, the wing pitches down from zero to maximal angle of attack (Figure \ref{img_motionParametrization}). The pitching velocity starts positive because the pitching rotation starts at the end of the upstroke, and reaches zero at the end of phase I.
It follows that the pitching acceleration $\ddot{\alpha}$ starts and ends at zero with a negative peak at $t'\approx0.01$. At the same time, the wing flaps from rest to its cruise translational velocity such that the acceleration $\ddot{\phi}$ decays from its minimum value to zero at $t'\approx 0.08$. 
This fast wing motion generates two successive lift peaks before decreasing to phase II (Figure \ref{img_liftKphiVar} top). The first peak is present for the three kinematics and its magnitude increases with both $K_\phi$ and $K_\alpha$. 
Low $K_\phi$ mostly influences the mean lift but conserves the force dynamics, unlike low $K_\alpha$ which preserves a similar mean lift but filters out the second peak. 
It is also interesting to note that the same trend occurs both in the first and in the fifth cycle. This implies that the wing-wake interaction is not the sole responsible for the observed lift trends.

\paragraph{Added mass}


For many species (drosophila \cite{Sane2001}, mosquitoes \cite{Liu2020}, hawkmoths\cite{Lua2010}, etc), the virtual mass effect is known to produce a peak in the lift at the beginning of the stroke. The peak is also observed on the pressure and the suction sides in the first and the fifth cycles of the investigated test case (marker (b) at $t'=0.01$ in Figure \ref{img_liftSide} left).
The lift peaks simultaneously with the vertical component of the flapping acceleration projected normally to the wing $\ddot{\phi}\sin\alpha\cos\alpha$ (see Figure \ref{img_motionParametrization}) and is also always larger on the suction side. Figure \ref{img_addM_pressureField} shows the pressure coefficient distribution on both sides of the wings at instants $t'\in[0.005,0.05]$.
Initially, a low-pressure ellipsoid occupies most of the suction side of the wing. It has a larger magnitude compared to the high pressures forming a "horseshoe" pattern around a half, low pressure ellipsoid on the pressure side. After $t'=0.01$, the ellipsoid shape of the suction side contains less negative pressure and moves towards the TE, decreasing the lift force. The larger pressure on the suction side differs from the observations of \citet{Van2022} who found that the added mass effect is primarily generated on the pressure side for various wings accelerating at a constant angle of attack.

\begin{figure}[!ht]\center
	 \includegraphics[width=0.45\textwidth]{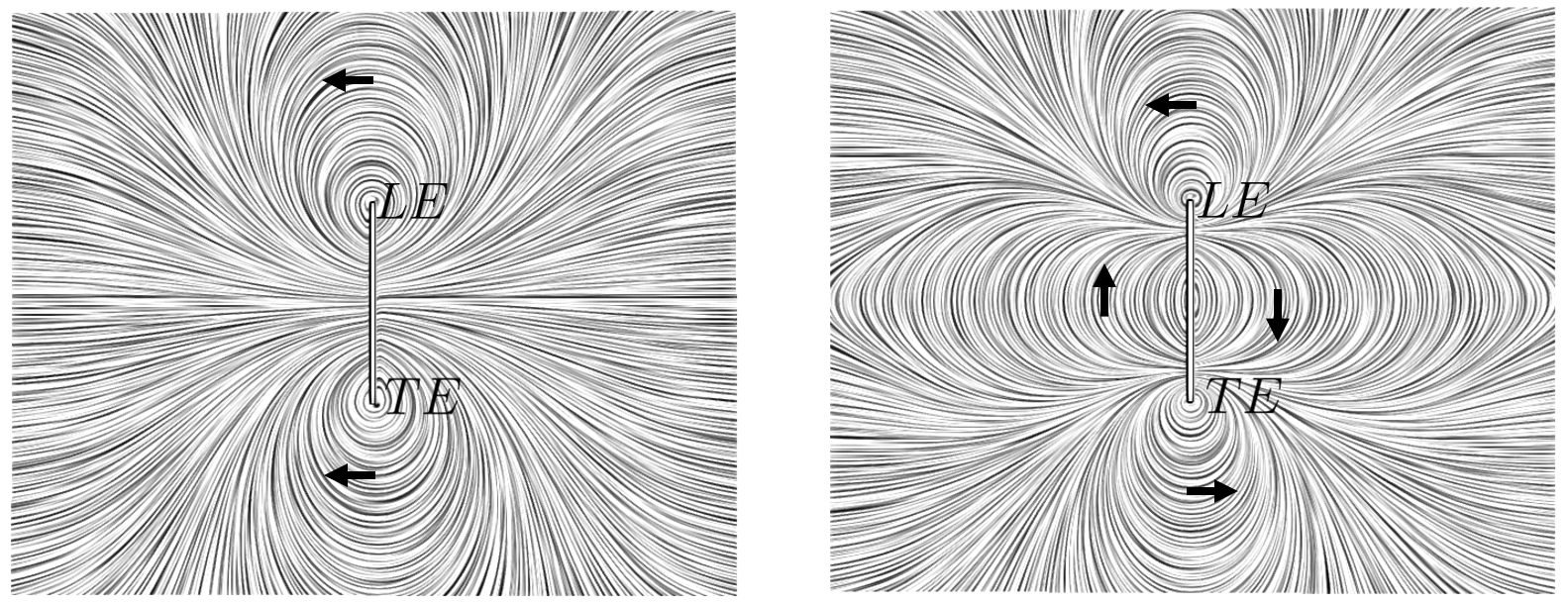}
     \caption{Velocity streamlines of a flapping motion (left) and a pitching motion (right) from a potential flow at $t'=0.02$.}
 \label{img_addM_pitchandFlap}    
\end{figure}

\begin{figure*}[!ht]\center
\begin{subfigure}{\textwidth}
    \begin{minipage}{0.05\textwidth} 
        \captionsetup{justification=raggedright,singlelinecheck=false,format=hang}
        \caption{}
        \centering
        \includegraphics[width=0.99\textwidth]{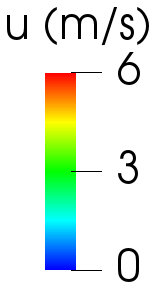}
        \label{img_fieldVel}
    \end{minipage}%
    \begin{minipage}{0.95\textwidth} 
        \centering
        \includegraphics[width=0.99\textwidth]{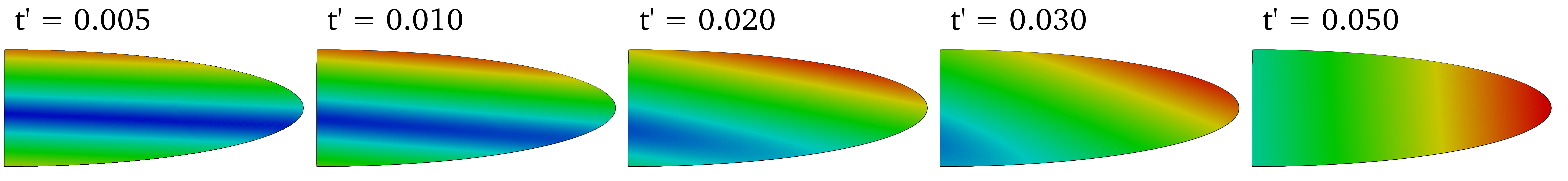}
    \end{minipage}
\end{subfigure}
\begin{subfigure}{\textwidth}
    \begin{minipage}{0.05\textwidth} 
        \captionsetup{justification=raggedright,singlelinecheck=false,format=hang}
        \caption{}
        \centering
        \includegraphics[width=0.99\textwidth]{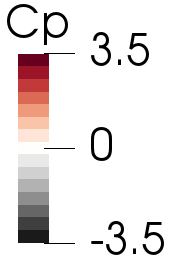}
        \label{img_pSuction_pot}
    \end{minipage}%
    \begin{minipage}{0.95\textwidth} 
        \centering
        \includegraphics[width=\textwidth]{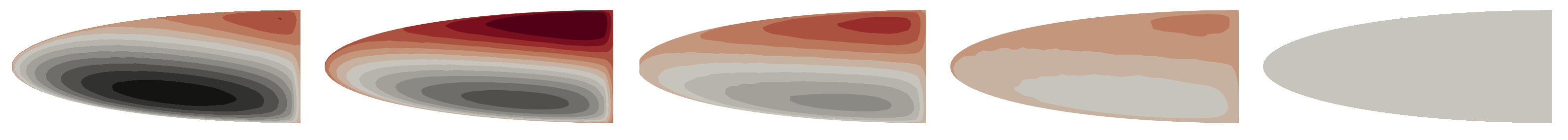}
    \end{minipage}
\end{subfigure}
\begin{subfigure}{\textwidth}
    \begin{minipage}{0.05\textwidth} 
        \captionsetup{justification=raggedright,singlelinecheck=false,format=hang}
        \caption{}
        \centering
        \includegraphics[width=0.99\textwidth]{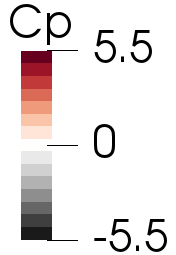}
        \label{img_pPressure_pot}
    \end{minipage}%
    \begin{minipage}{0.95\textwidth} 
        \centering
        \includegraphics[width=0.99\textwidth]{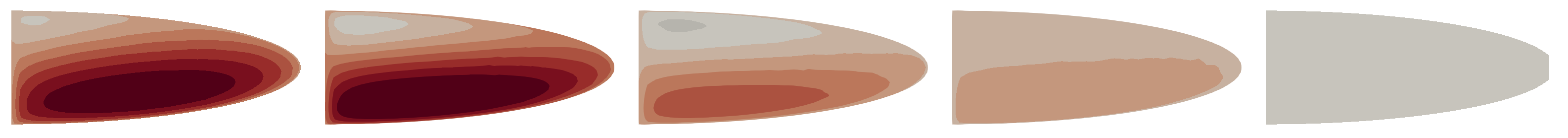}
    \end{minipage}
\end{subfigure}
\begin{subfigure}{\textwidth}
    \begin{minipage}{0.05\textwidth} 
        \captionsetup{justification=raggedright,singlelinecheck=false,format=hang}
        \caption{}
        \centering
        \includegraphics[width=0.99\textwidth]{./legendU_2.png}
        \label{img_fieldAddM1}
    \end{minipage}%
    \begin{minipage}{0.95\textwidth} 
        \centering
        \includegraphics[width=0.99\textwidth]{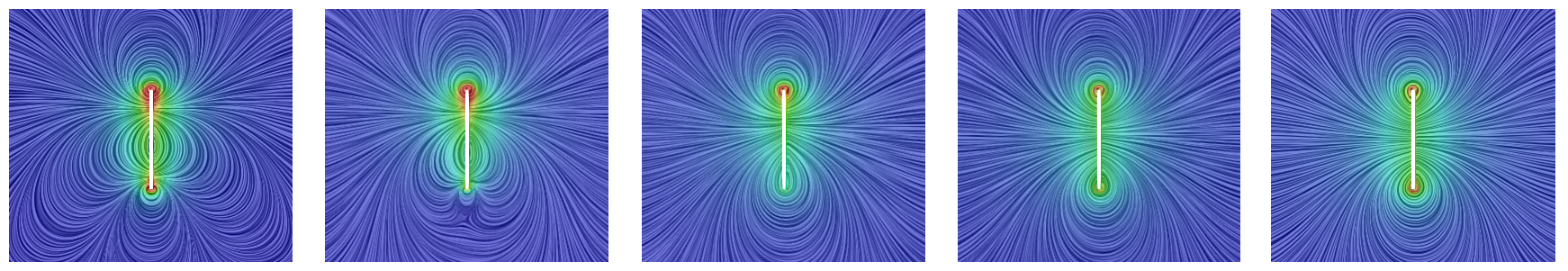}
    \end{minipage}
\end{subfigure}
\begin{subfigure}{\textwidth}
    \begin{minipage}{0.05\textwidth} 
        \vspace{-0.4cm}
        \captionsetup{justification=raggedright,singlelinecheck=false,format=hang}
        \caption{}
        \centering
        \includegraphics[width=0.99\textwidth]{./legendU_2.png}
        \label{img_fieldAddM2}
    \end{minipage}%
    \begin{minipage}{0.95\textwidth} 
        \centering
        \includegraphics[width=0.99\textwidth]{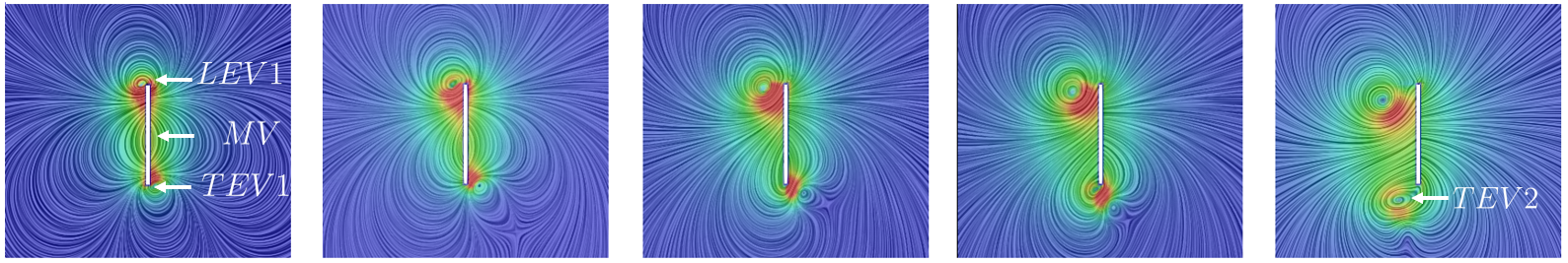}
    \end{minipage}
\end{subfigure}
     \caption{(a) Velocity magnitude of the wing used to simulate the potential flow resulting in pressure distributions on (b) the suction side and (c) pressure side. The streamlines and velocity field are also shown on the $R_2$ slice in (d) potential flow and (e) viscous flow for the 1st cycle at instants corresponding to the first five markers of Figure \ref{img_liftSide}.}
 \label{img_fieldAddM}    
\end{figure*}

The pressure distributions of Figure \ref{img_addM_pressureField} are associated with the added mass effect by using potential flow computations (section \ref{sec_potentialFlow}). Figure \ref{img_addM_pitchandFlap} first shows the streamlines of two separate potential flows, one with flapping and one with pitching boundary conditions at $(K_\phi,K_\alpha)=(0.99,10)$.
The results are shown on a plane at span $x_\alpha=R_2$, with $z_\alpha$ oriented vertically for $t'=0.02$. This plane, named $R_2$ slice, is always used for the other figures unless specified differently.  
The left figure (flapping motion) allows to identify a doublet flow and a transversal flow. The two recirculation zones, anti-clockwise at the LE and clockwise at the TE,  expand from the airfoil tip and have a teardrop shape due to the action of the transverse flow. The latter is perpendicular to the chord as visible close to the wing center. 
In the right figure (pitching motion), two anti-clockwise vortices are separated by a region of counter-rotating vortex. These flow patterns correspond to the ones found using vortex sheet theory and experiments in \citet{Corkery2019}.

The two velocity potentials are then added together to obtain the potential flow of the flapping and pitching wing (Figure \ref{img_fieldAddM}). Figure \ref{img_fieldVel} shows the velocity field on the wing surface, Figure \ref{img_pSuction_pot} and \ref{img_pPressure_pot} show the pressure coefficient distribution on the suction and pressure surface and Figure \ref{img_fieldAddM1} compares the streamlines and velocity magnitude of the potential flow with the viscous flow from the first cycle (Figure \ref{img_fieldAddM2}) for instants in $t'\in[0.005,0.05]$ that correspond to markers (a) to (e) in Figure \ref{img_liftSide}.
Obvious similarities are visible between the pressure distributions on the suction side of the wing for the potential and viscous flow. A similar ellipsoid of low pressure is moving towards the TE and the root of the wing, following a mid-chord vortex shown with the streamlines in Figure \ref{img_fieldAddM1}.
The pressure distribution on the pressure side (Figure \ref{img_pPressure_pot})  bears less resemblance to the viscous distribution (Figure \ref{img_ppressure_1st}). It shows a high pressure region attached to the TE but does not reproduce the horseshoe pattern. 

Both potential and viscous computations show three bounded recirculation zones at the LE (LEV1, anti-clockwise), mid-chord (MV, clockwise) and TE (TEV1, anti-clockwise) and a zone of transverse flow between the LE and the mid-chord. At $t'=0.005$, the streamlines are very similar to the pitching-only case (Figure \ref{img_addM_pitchandFlap} right) with velocity magnitudes that are larger at the LE than at the TE because the flapping velocities add to the pitch-down velocities.
This results in a larger LE vortex (Figure \ref{img_fieldAddM1} and \ref{img_fieldAddM2}) that grows and detaches when viscous effects are accounted for.  


The MV initially builds around the pitching axis where velocities are minimal at $t'\sim0.005$. This zone moves towards the TE when the pitching angle increases, and replaces TEV1 which is shed in the viscous flow ($t'=0.01$).
The newly formed TE vortex (clockwise) preserves its size in the potential flow  ($t'=[0.03,0.05]$) and depends only on the instantaneous flapping motion at $t'=0.05$ (Figure \ref{img_fieldAddM1}). In the viscous flow, the new TEV (TEV2) grows and eventually sheds. As one could expect, viscous effects become more pronounced with the decrease of the wing acceleration. 



\begin{figure}[!htb]\center
	 \includegraphics[width=0.48\textwidth]{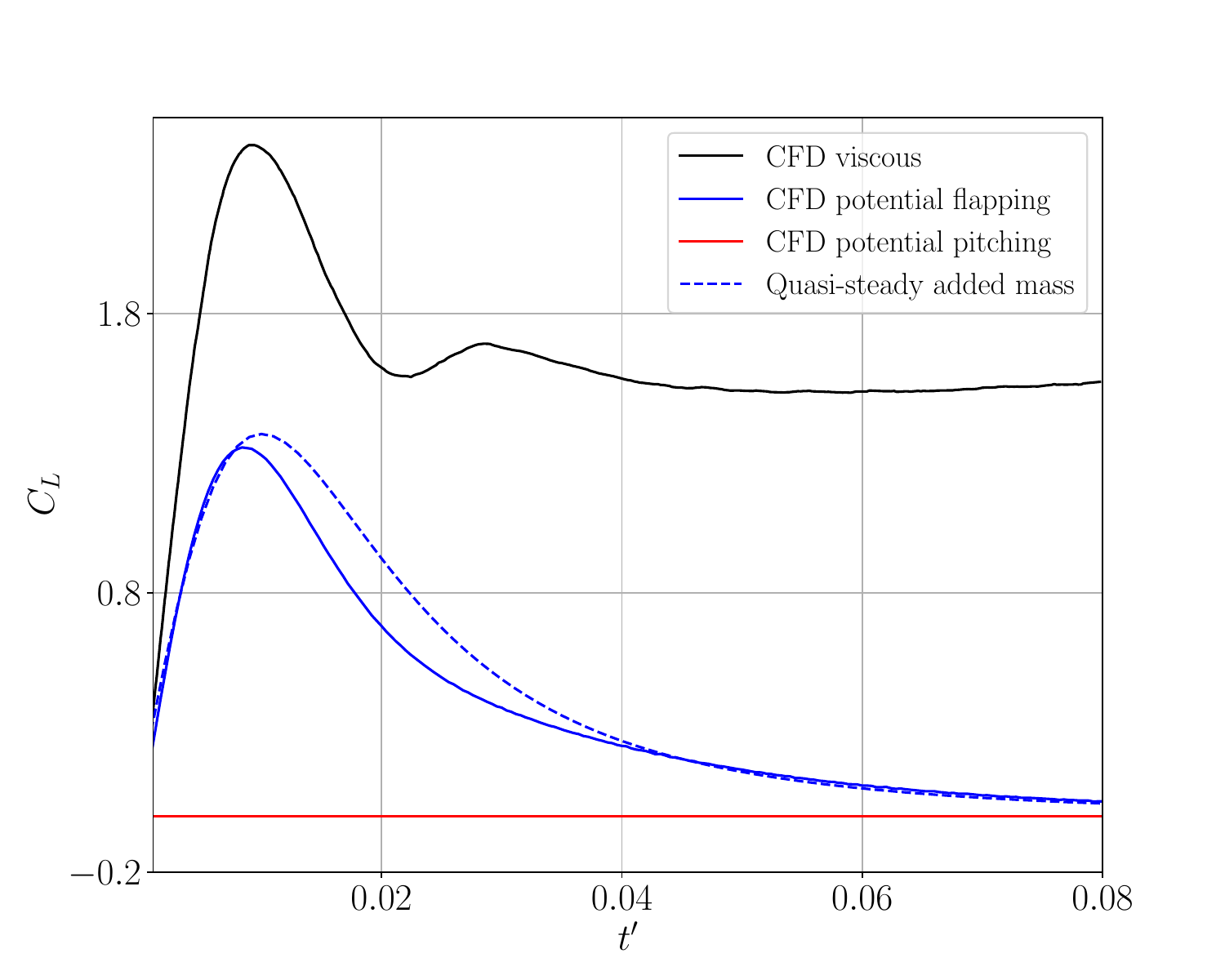}
     \caption{Lift computed with CFD computations and compared with a quasi-steady model \cite{Lee2016} for the initial stages of the first flapping cycle.}
 \label{img_LiftAddM}    
\end{figure}

The same differences are seen in the lift evolution in Figure \ref{img_LiftAddM}, which compares the potential flow, the viscous flow, and a quasi-steady model by \citet{Lee2016}. The latter considers that only a cylindrical volume of air with a radius equal to the chord is accelerated.
Overall, the potential flow and the quasi-steady models predict lower lift. This is different than observed by \citet{Liu2020} who attributes the lift reduction in the viscous flow to the wing-wake interaction. 
Both the potential flow and the quasi-steady model capture the initial peak confirming the role of the virtual mass effect in the aerodynamic loads at the beginning of the stroke.

Figure \ref{img_LiftAddM} also shows that the initial peak is solely generated by the flapping motion as a consequence of the potential flow symmetry and the mid-chord position of the pitching axis. The pitching rotation generates opposing reaction forces on each side of the wing so that no net force is produced. This seems contradictory to Figure \ref{img_liftKphiVar} which shows that highly dynamic pitching kinematics increases the first lift peak. Nevertheless, this results only from the larger angles of attack that redistribute more of the "flapping" added mass in the lift direction and less in the drag direction. 
Hence, potential flow effects do not explain the second peak that arises when $K_\alpha$ increases (see Figure \ref{img_liftKphiVar}). It can also be seen that this second peak is not linked to vortical events (LEV detachment, etc). The rest of the section analyses the second peak's origin.

For $t'\in [0.005,0.02]$, the lift decreases on the pressure side of the wing (Figure \ref{img_liftSide}, left) as confirmed by the decreasing mean value on the surface pressure field shown in Figure \ref{img_ppressure_1st}. 
As the pitching velocity decreases, the high-pressure region on the LE is replaced by the low-pressure region formed at the center of the wing root. The latter increases in magnitude and expands radially and along the chord. TEV1 is simultaneously shed, but this has negligible influence on the pressure distribution. 
The described trend for $t'\in[0.005,0.02]$ results from the highly dynamic pitching-down motion. 



\begin{figure*}[!ht]\center
\begin{subfigure}{0.47\textwidth}
    \begin{minipage}{0.05\textwidth} 
        \vspace{-2cm}
        \captionsetup{justification=raggedright,singlelinecheck=false,format=hang}
        \caption{}
        \label{img_ftleStart_0}
    \end{minipage}%
    \begin{minipage}{0.95\textwidth} 
        \centering
        \includegraphics[width=0.99\textwidth]{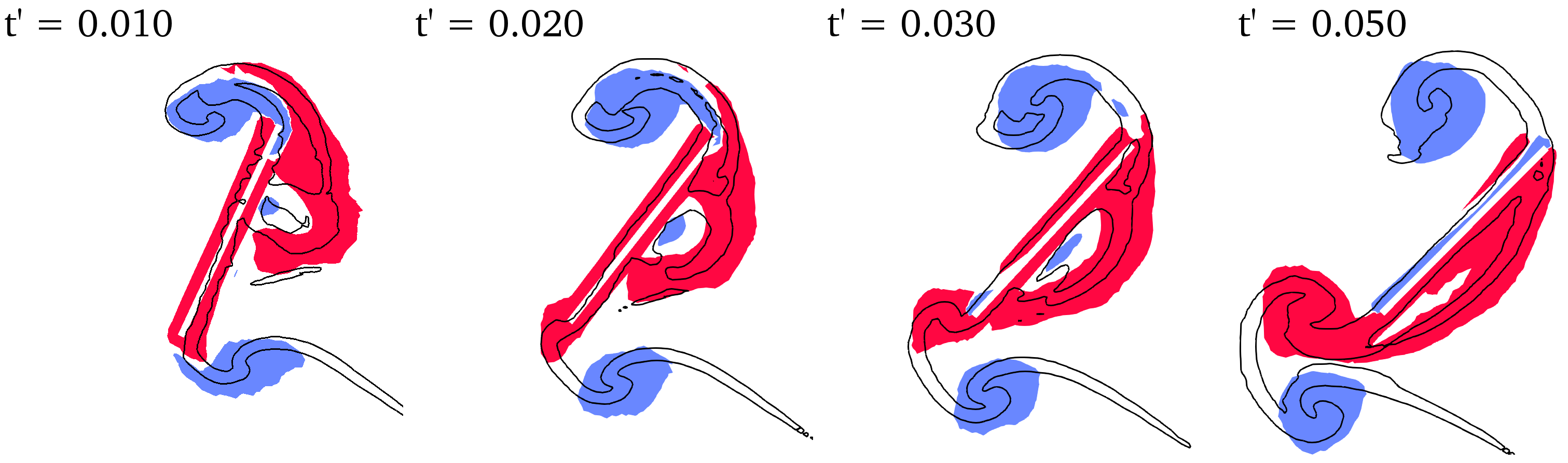}
    \end{minipage}
\end{subfigure}
\begin{subfigure}{0.47\textwidth}
    \begin{minipage}{0.05\textwidth} 
        \vspace{-2cm}
        \captionsetup{justification=raggedright,singlelinecheck=false,format=hang}
        \caption{}
        \label{img_ftleStart_1}
    \end{minipage}%
    \begin{minipage}{0.95\textwidth} 
        \centering
        \includegraphics[width=0.99\textwidth]{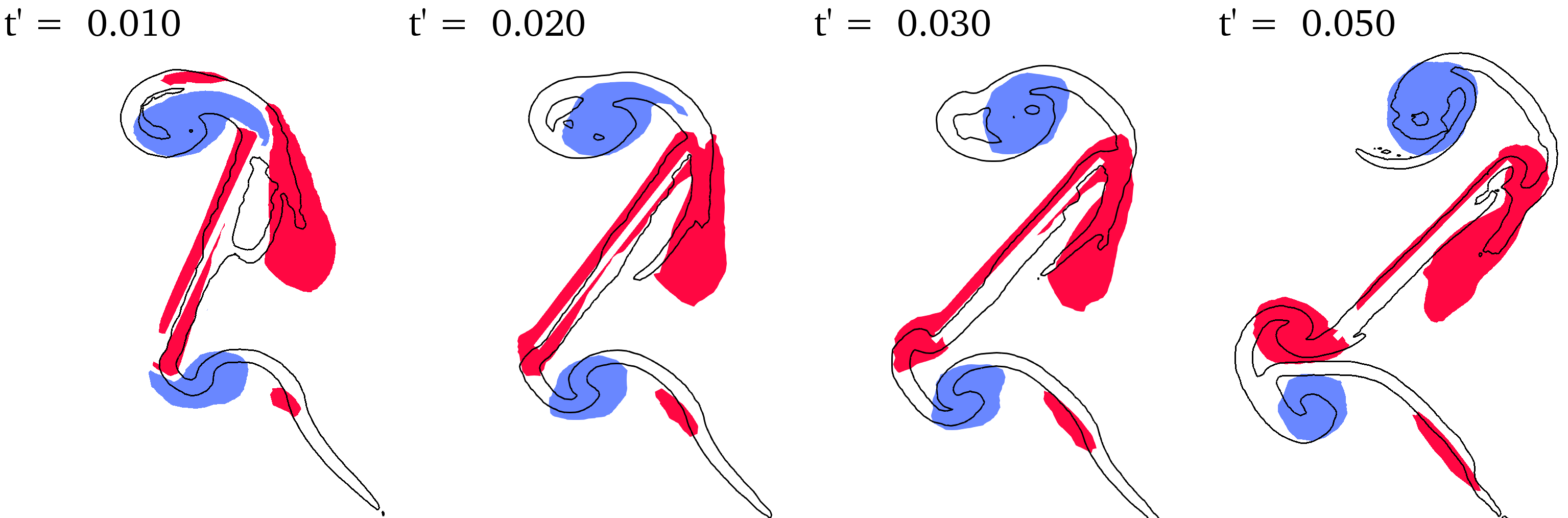}
    \end{minipage}
\end{subfigure}
\begin{subfigure}{0.45\textwidth}
    \begin{minipage}{0.05\textwidth} 
        \vspace{-2cm}
        \captionsetup{justification=raggedright,singlelinecheck=false,format=hang}
        \caption{}
        \label{img_ftleStart_2}
    \end{minipage}%
    \begin{minipage}{0.95\textwidth} 
        \centering
        \includegraphics[width=0.99\textwidth]{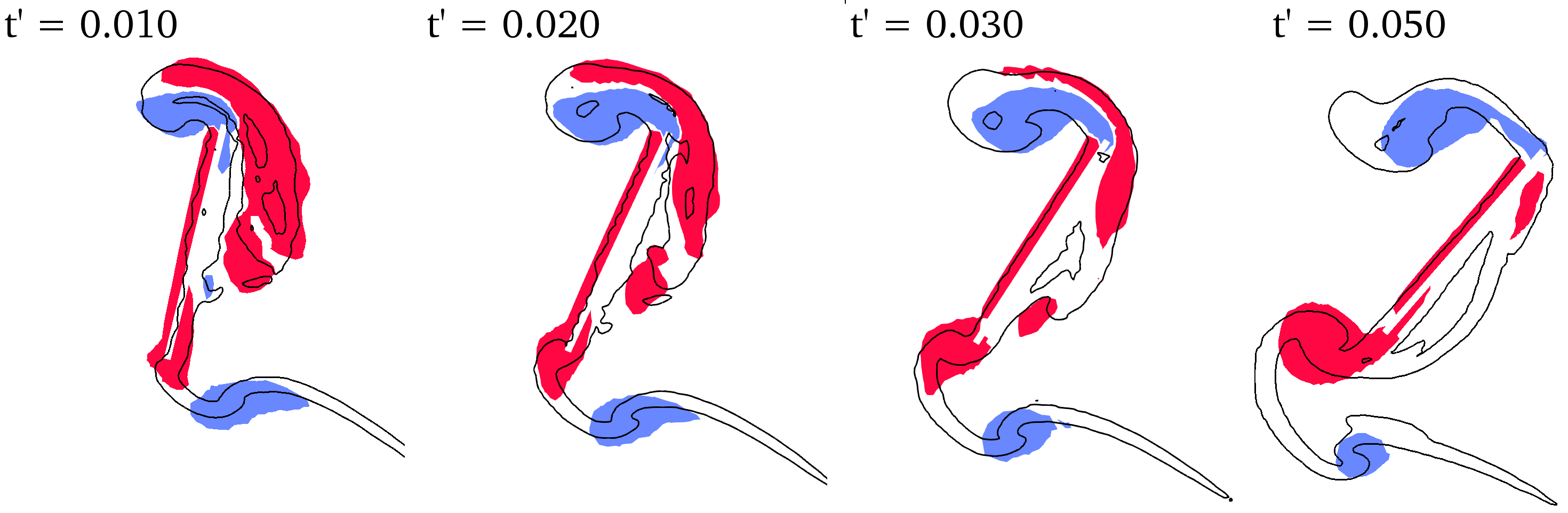}
    \end{minipage}
\end{subfigure}
     \caption{Starting vortices during phase I, highlighted by a contour of the FTLE field and the spanwise vorticity field $\omega_{x_\alpha}$ at a radial plane $x_\alpha=R_2$ for (a) $(K_\phi,K_\alpha)=(0.99,10)$, (b) $(K_\phi,K_\alpha)=(0.5,10)$ and, (c) $(K_\phi,K_\alpha)=(0.99,5)$.}
 \label{img_ftleStart}    
\end{figure*}

\begin{figure}[!ht]\center
	 \includegraphics[width=0.49\textwidth]{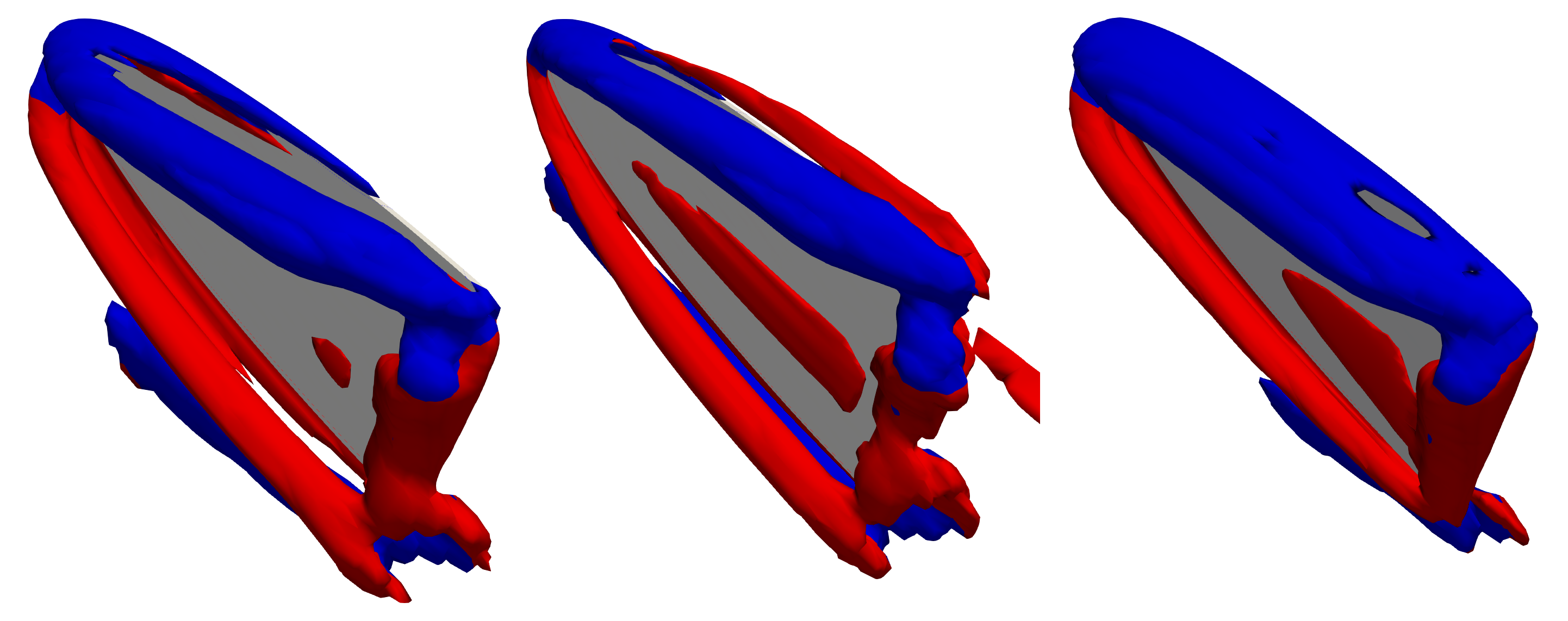}
     \caption{Starting vortices during phase I at $t'=0.05$, highlighted by the contour of the Q-criterion and colored with $\omega_{x_\alpha}$. Left: $(K_\phi,K_\alpha)=(0.99,10)$, middle: $(K_\phi,K_\alpha)=(0.5,10)$  and right: $(K_\phi,K_\alpha)=(0.99,5)$.}
 \label{img_wyStart}    
\end{figure}

When $t'>0.02$, the influence of the flapping motion becomes dominant again and the lowest pressures at the LE (Figure \ref{img_ppressure_1st}) are replaced with higher pressures that expand from the tip. This increases the lift force (Figure \ref{img_liftSide}, left). After $t'\sim0.03$, the high-pressure region moves towards the tip, the lift decreases on the pressure side and a second peak is visible on the pressure side. 

When combined with the lift from the suction side, this gives a well-defined second peak of the lift for high $K_\alpha$.
For low $K_\alpha$, there is no second peak during the first downstroke. This can be explained by the longer descending and ascending phases of the pressure lift that are more in phase with the suction lift. 



\vspace{0.2cm}
\paragraph{Vortex Dynamics} 

Figure \ref{img_fieldAddM2} showed that the high pitching velocity of the wing forces the flow to separate at the LE and TE of the wing. The flow rolls up into two anti-clockwise vortices, a leading-edge vortex (LEV1) and a trailing-edge vortex (TEV1), that form above the upper surface and lower surface respectively. Figure \ref{img_ftleStart} illustrates the evolution of the unbound LEV1 and TEV1 with the spanwise vorticity $\omega_{x_\alpha}$, where blue colors indicate the anti-clockwise vortices and red clockwise vortices. The contours of the FTLE field ($\bm{\sigma}/\bm{\sigma}_{max} = 0.2$) are overlapped in black on the figure which focuses on the 5th cycle.
For $(K_\phi,K_\alpha)=(0.99,10)$ in Figure \ref{img_ftleStart_0}, both LE and TE vortices rapidly grow, reach a maximum size and detach due to the high pitching velocity that prevents reattachment.
LEV1 remains initially closely above the LE as it is linked to it through a vortex filament ($t'\sim 0.02$). The vortex filament then breaks and the LEV sheds from the LE to the TE to reach the wake in phase II. 
TEV1 also rapidly detaches and goes straight in the wake as the wing passes over it. Shortly after its detachment, the second clockwise TEV2 forms and shed together with TEV1 ($t'\sim 0.05$).
Figure \ref{img_wyStart} shows the 3D vortical structures with the Q-criterion field. The figure considers the first downstroke to better render the vortical structures generated at the leading edge, the trailing edge, and the root of the wing. TEV2 and LEV1 connect at the tip and root of the wing to form a loop vortex. The loop eventually distorts and breaks at the root as vortices are shed in the wake at different velocities.

The wing motion dynamics modifies only marginally the vortex dynamics. 
In the early stage, a more dynamic pitching motion results in slightly larger LEVs and TEVs that are further from the wing (Figure \ref{img_ftleStart_0} and \ref{img_ftleStart_2}). 
The influence of the flapping dynamics is more visible when the minimum angle of attack is reached. LEV1, TEV1, and TEV2 shed later when $K_{\phi}$ decreases, owing to the lower instantaneous flapping velocity $\dot{\phi}$.
Interestingly, as $K_{\phi}$ decreases, a second LEV builds on the pressure side of the wing (Figure \ref{img_ftleStart_1} and \ref{img_wyStart}). It rotates clockwise with a comparable magnitude as TEV2.
Overall, the different vortical structures are small and detached from the wing during most of phase I. Their associated pressure drops have a minor contribution to the wing load.

In Figure \ref{img_ftleStart}, the wing also meets an anti-clockwise vorticity region from the LEV of the previous upstroke. The following of this section correlates this wing-wake interaction mechanism with the lift.

\begin{figure*}[!ht]\center
\begin{subfigure}{\textwidth}
    \begin{minipage}{0.05\textwidth} 
        \vspace{-0.3cm}
        \captionsetup{justification=raggedright,singlelinecheck=false,format=hang}
        \caption{}
                \centering
        \includegraphics[width=\textwidth]{./legend_psuction_1stCycle_2.png}
        \label{img_wake1}
    \end{minipage}%
    \begin{minipage}{0.95\textwidth} 
        \centering
        \includegraphics[width=\textwidth]{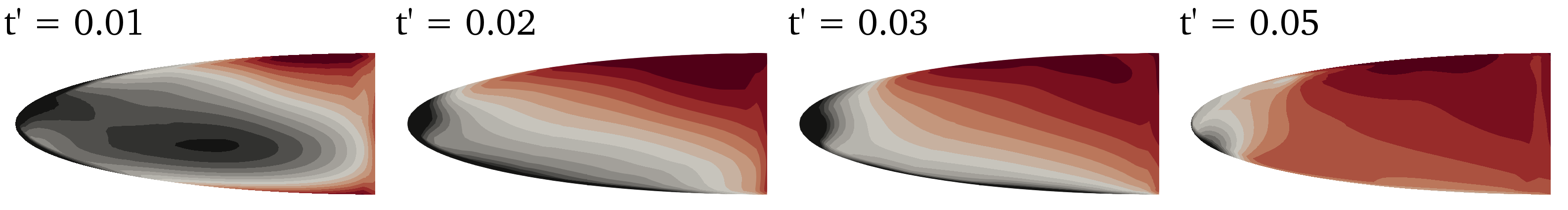}
    \end{minipage}
\end{subfigure}
\begin{subfigure}{\textwidth}
    \begin{minipage}{0.05\textwidth} 
        \vspace{-0.4cm}
        \captionsetup{justification=raggedright,singlelinecheck=false,format=hang}
        \caption{}
                        \centering
        \includegraphics[width=\textwidth]{./legend_p_1stCycle_2.png}
        \label{img_wake2}
    \end{minipage}%
    \begin{minipage}{0.95\textwidth} 
        \centering
        \includegraphics[width=\textwidth]{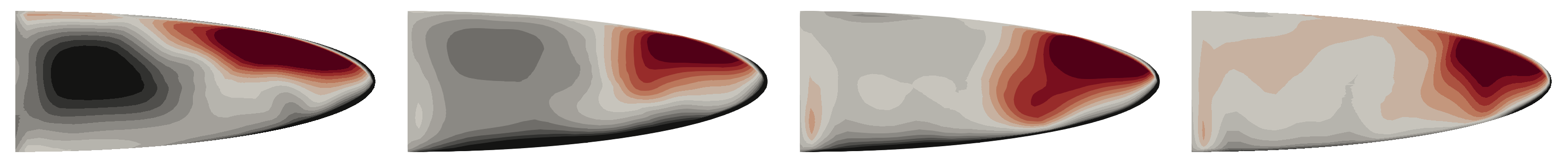}
    \end{minipage}
\end{subfigure}
\begin{subfigure}{\textwidth}
    \begin{minipage}{0.05\textwidth} 
        \vspace{-2.5cm}
        \captionsetup{justification=raggedright,singlelinecheck=false,format=hang}
        \caption{}
                \centering
        \includegraphics[width=\textwidth]{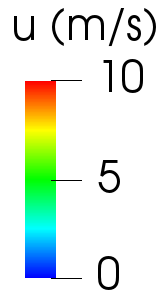}
        \label{img_wake3}
    \end{minipage}%
    \begin{minipage}{0.95\textwidth} 
        \centering
        \includegraphics[width=\textwidth]{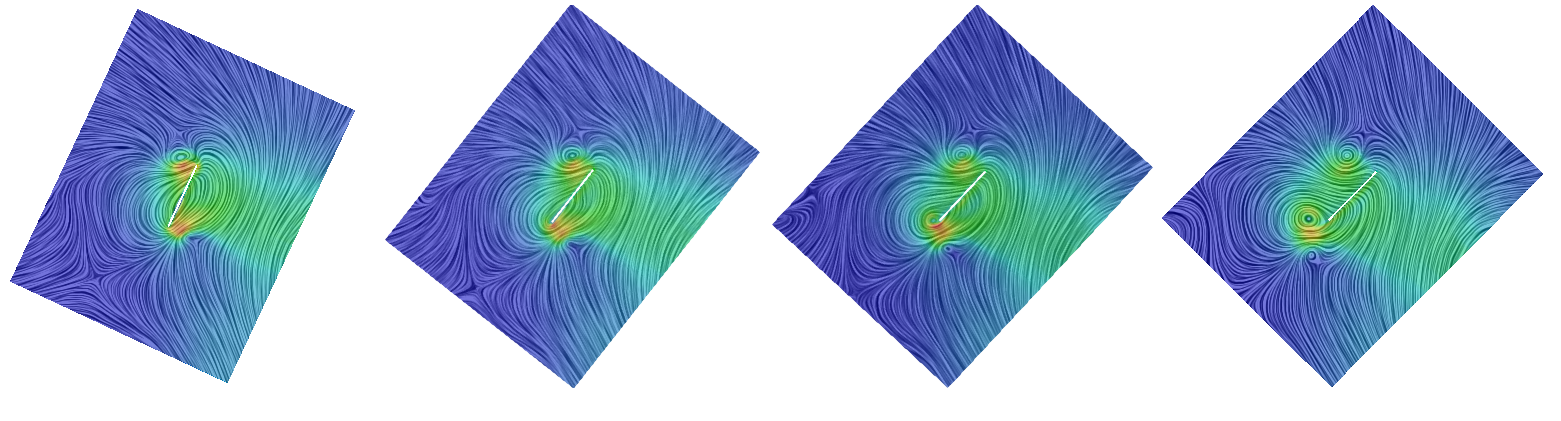}
    \end{minipage}
\end{subfigure}
\begin{subfigure}{0.48\textwidth}
    \begin{minipage}{0.05\textwidth} 
        \vspace{-3.7cm}
        \captionsetup{justification=raggedright,singlelinecheck=false,format=hang}
        \caption{}
        \label{img_wake4}
    \end{minipage}%
    \begin{minipage}{0.95\textwidth} 
        \centering
        \includegraphics[width=\textwidth]{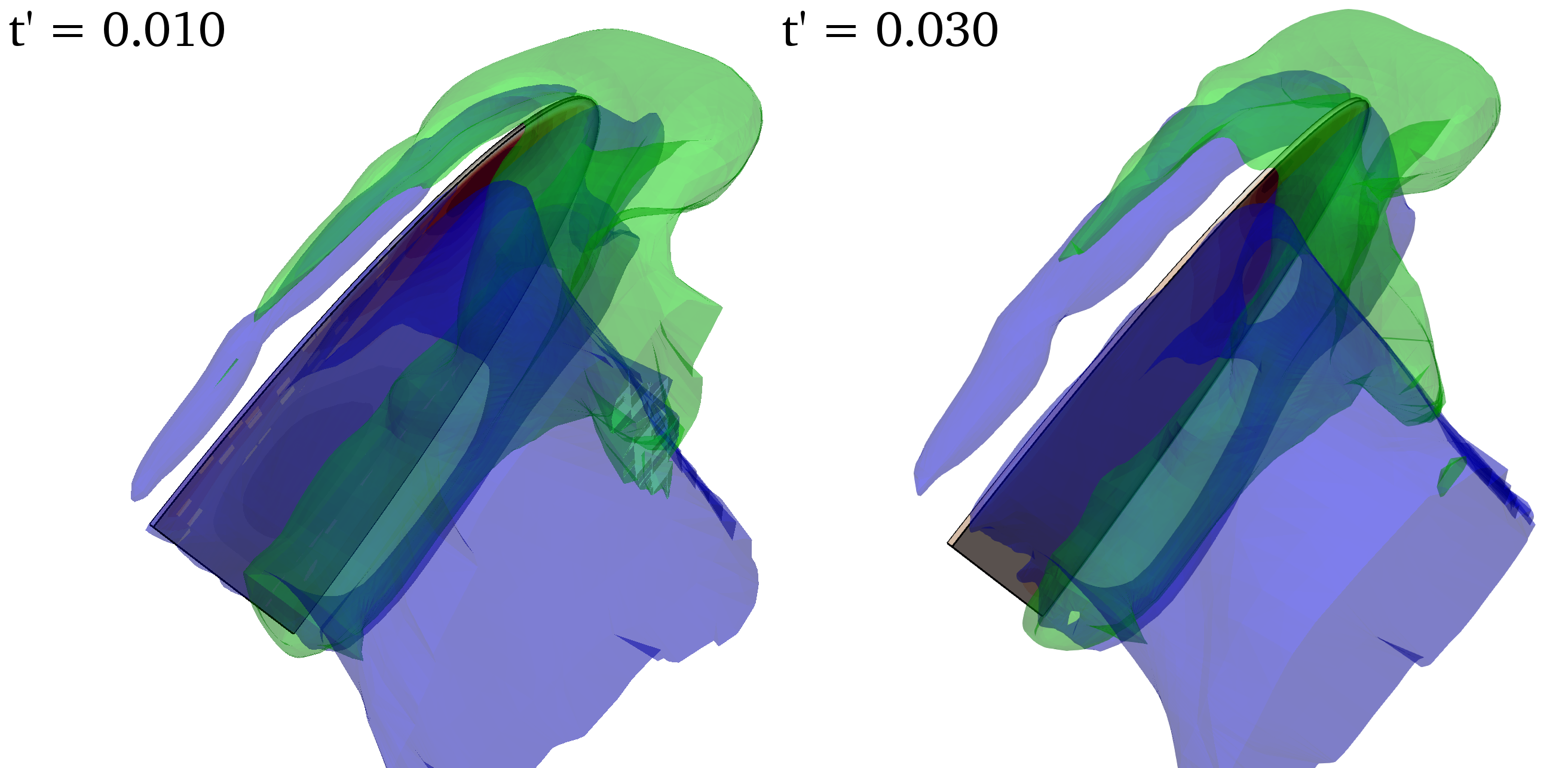}
    \end{minipage}
\end{subfigure}
\begin{subfigure}{0.48\textwidth}
    \begin{minipage}{0.05\textwidth} 
        \vspace{-3.7cm}
        \captionsetup{justification=raggedright,singlelinecheck=false,format=hang}
        \caption{}
        \label{img_wake5}
    \end{minipage}%
    \begin{minipage}{0.95\textwidth} 
        \centering
        \includegraphics[width=\textwidth]{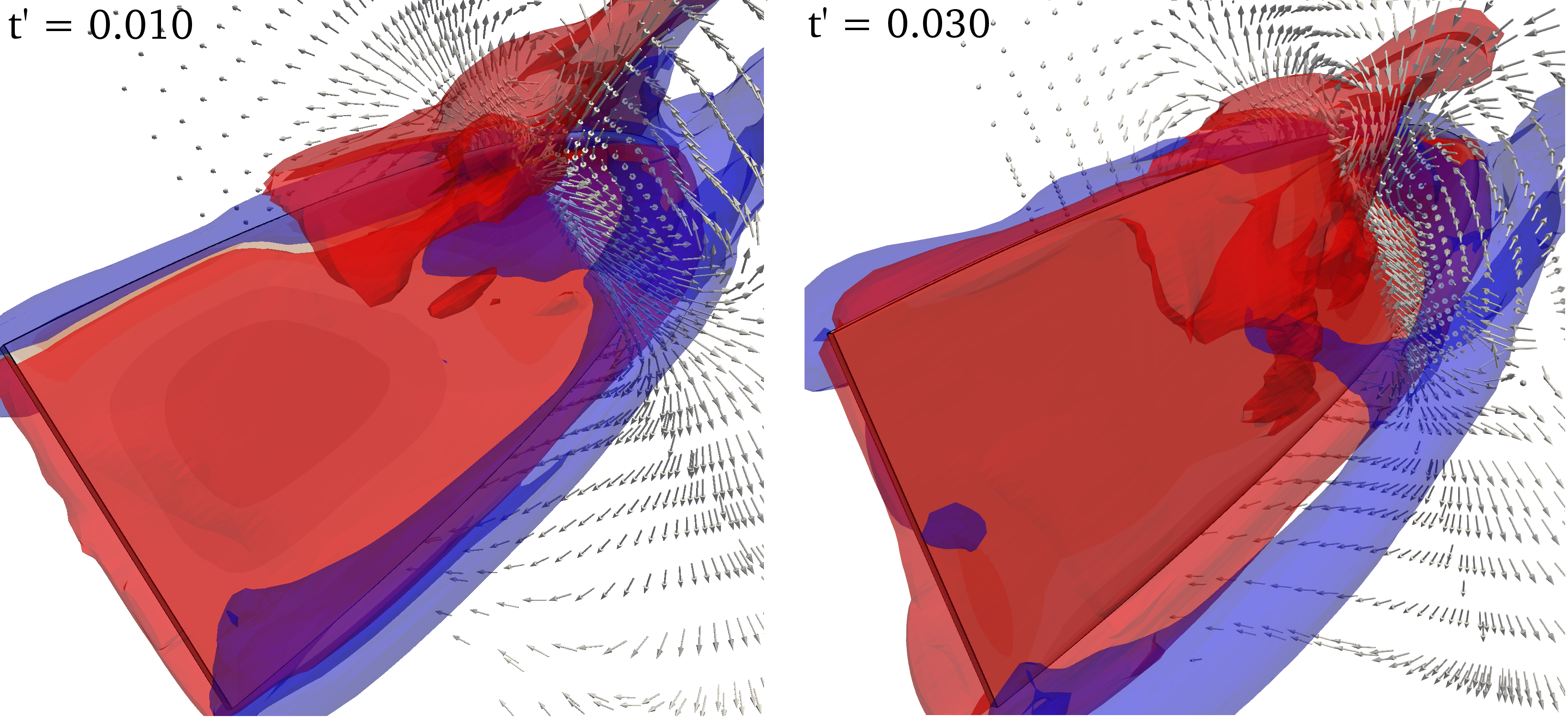}
    \end{minipage}
\end{subfigure}
     \caption{Influence of the wing-wake interaction on the pressure distribution of (a) the suction side, (b) the pressure side as well as on (c) the velocity field and streamlines of the $R_2$ slice for $t'\in[0.01,0.05]$ evidenced with markers in Figure \ref{img_liftSide}. (d) shows volumes that satisfies $u_{z_I}<-2$ (blue) and  $u_{y_\phi}<$-2 for $t'\in[0.01,0.03]$. (e) shows the radial vorticity and the velocity vectors at a slice $x_\alpha\sim1.05R_2$.}
 \label{img_ww}    
\end{figure*}

\vspace{0.2cm}
\paragraph{Wing-wake interaction}
Figure \ref{img_liftKphiVar} clearly shows that the wake from the previous strokes changes the time evolution of the lift, decreasing its mean value. 
On the suction side, the lift increases as shown with the pressure distributions in Figure \ref{img_wake1}. The wake velocity increases the air inertia, resulting in a larger and stronger low-pressure ellipsoid characteristic of the added mass (Figure \ref{img_wake1}, $t'=0.01$).
The recirculation region MV occupies a larger portion of the chord since it is generated by the pitching up (end of 4th cycle) and the pitching down (start of the 5th cycle). 
The complete pitching motion also increases the size and the strength of TEV1 and LEV1 which shed earlier as seen with the streamlines and velocity magnitude in Figure \ref{img_wake3}. As a result, LEV1 increases the suction pressure applied to the wing close to its tip ($t'=0.01$ on Figure \ref{img_wake1}). 


On the pressure side, the wing-wake interaction translates into a negative lift peak that results from two wake mechanisms described in \citet{Lee2018}. Indeed, two pressure regions are seen on the wing surface (Figure \ref{img_wake2}) in comparison to the sole horseshoe pattern formed during the first cycle (Figure \ref{img_ppressure_1st}).
The proximal part of the wing is covered by a wide region of negative pressure that results from the downwash flow of the previous upstroke. This negative velocity region is depicted with a shaded blue color in Figure \ref{img_wake4}, where the volume satisfying $u_{z_I}<-2$ m/s is shown. The downwash flow aligns the normal streamlines with the wing, reducing the lifting pressure.

The distal part of the wing is occupied by a region of strong positive pressure (Figure \ref{img_wake2}). The region is (1) more concentrated towards the tip and LE, and (2) with higher pressure than during the first stroke (Figure \ref{img_ppressure_1st}).
The first difference can be explained by the suction of TEV1, which is stronger than during the first downstroke. 
The second difference can be linked to the highest velocities faced by the wing due to the upstroke wake. Figure \ref{img_wake4} proves this by showing in shaded green the volume with velocities $u_{y_\phi}<-2$ m/s. This region is actually the distal part of the LEV2, continuously shed at the wing's tip during the upstroke (see next section). 
Another interesting aspect of the wing-wake interaction is the formation of a vortex system at the tip. 
Figure \ref{img_wake5} shows the clockwise vorticity (red) and anti-clockwise vorticity (blue) together with the velocity vector on a slice at $x_\alpha\sim1.05R_2$. The clockwise vorticity region results from the LEV2 of the upstroke and the counter-clockwise vorticity is mostly contained in the TEV1 from the downstroke. They induce a jet towards the wing tip that increases the pressure on its surface. The impinging zone of the jet grows as the two vortices get closer to it and reach a maximum at $t'\approx 4.03$ (Figure \ref{img_wake5}). 

In conclusion, one can remark that the proximal pressure drop is the wing-wake interaction mechanism that dominates the pressure dynamics. It has greater importance than the distal pressure gain on the pressure side and the pressure drop on the suction side. It increases the peak-to-peak amplitude of the second lift spike during the 5th cycle for $K_\alpha =10$ and generates the second lift peak for $K_\alpha =5$ (Figure \ref{img_liftKphiVar}).





\subsection{Phase II: Cruise}


\paragraph{Vortex dynamics} 

\begin{figure*}[p]\center
\begin{subfigure}{\textwidth}
    \begin{minipage}{0.05\textwidth} 
        \vspace{-7.8cm}
        \captionsetup{justification=raggedright,singlelinecheck=false,format=hang}
        \caption{}
        \label{img_ftle_K09910}
    \end{minipage}%
    \begin{minipage}{0.9\textwidth} 
        \centering
        \includegraphics[width=\textwidth]{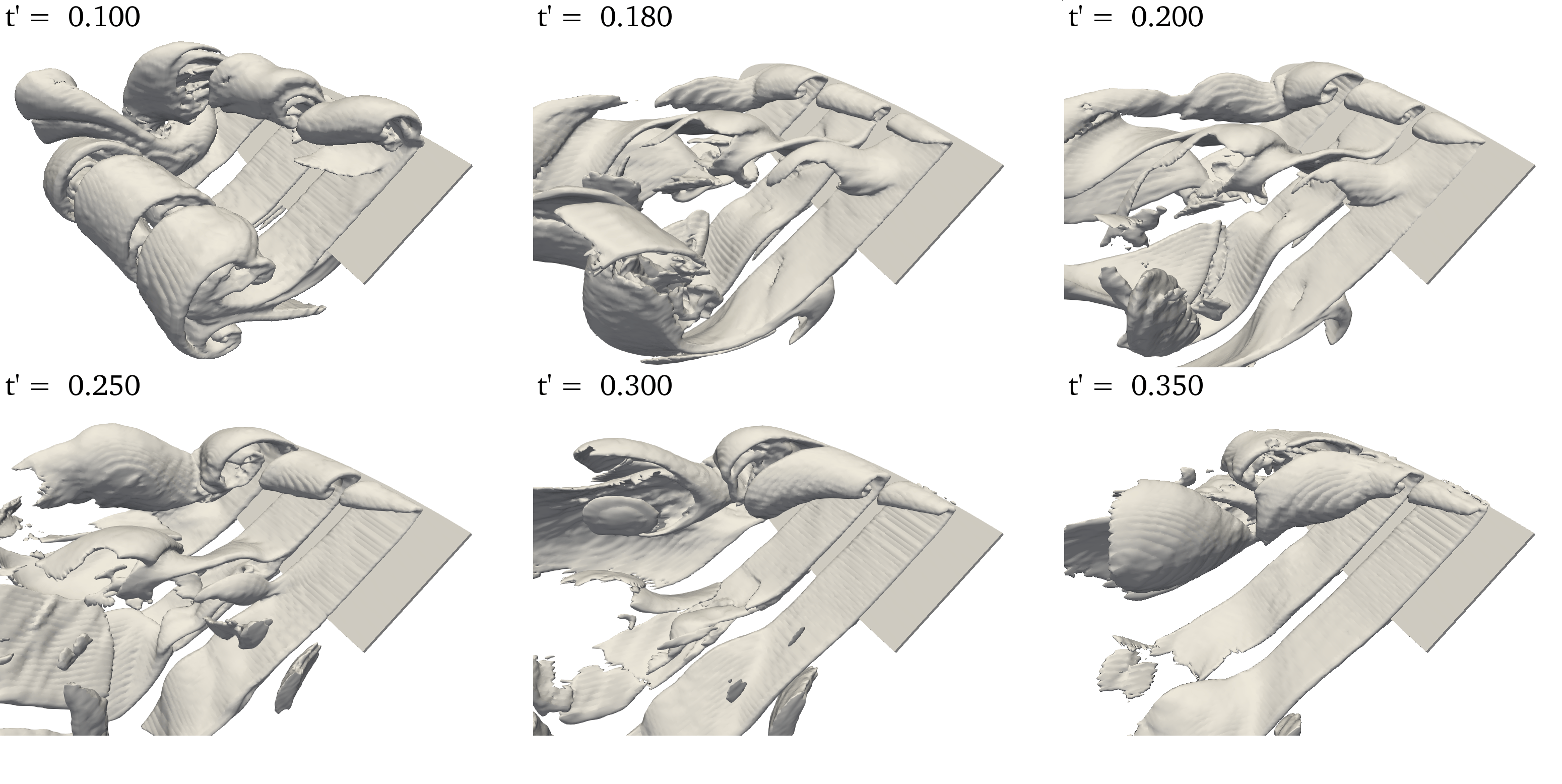}
    \end{minipage}
\end{subfigure}
\begin{subfigure}{\textwidth}
    \begin{minipage}{0.05\textwidth} 
        \vspace{-7.8cm}
        \captionsetup{justification=raggedright,singlelinecheck=false,format=hang}
        \caption{}
        \label{img_ftle_K0510}
    \end{minipage}%
    \begin{minipage}{0.9\textwidth} 
        \centering
        \includegraphics[width=\textwidth]{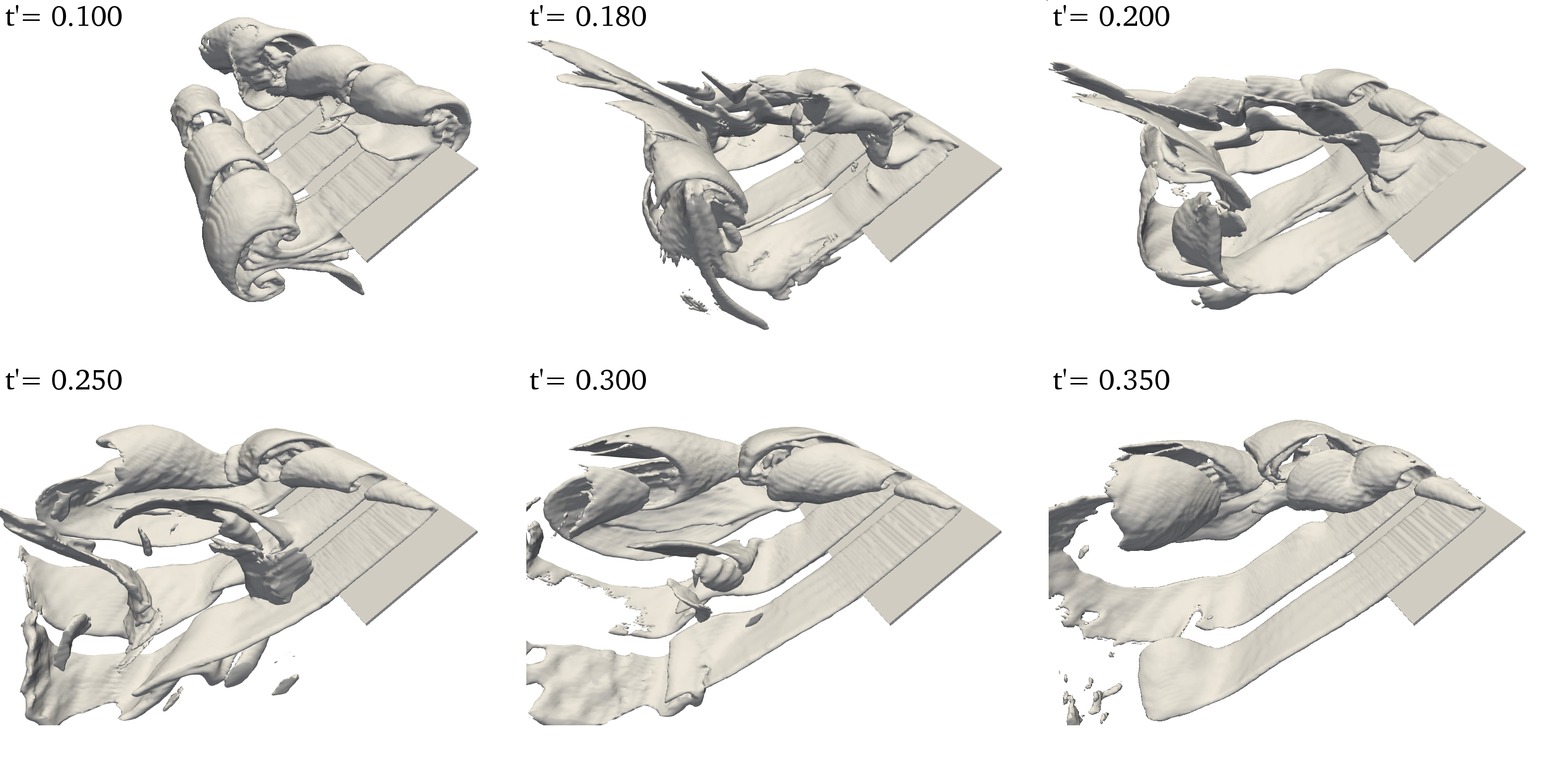}
    \end{minipage}
\end{subfigure}
\begin{subfigure}{\textwidth}
\begin{minipage}{0.65\textwidth}
    \begin{minipage}{0.05\textwidth} 
        \vspace{-5.1cm}
        \captionsetup{justification=raggedright,singlelinecheck=false,format=hang}
        \caption{}
        \label{img_ftle_K0995}
    \end{minipage}%
    \begin{minipage}{0.9\textwidth} 
        \centering
        \includegraphics[width=\textwidth]{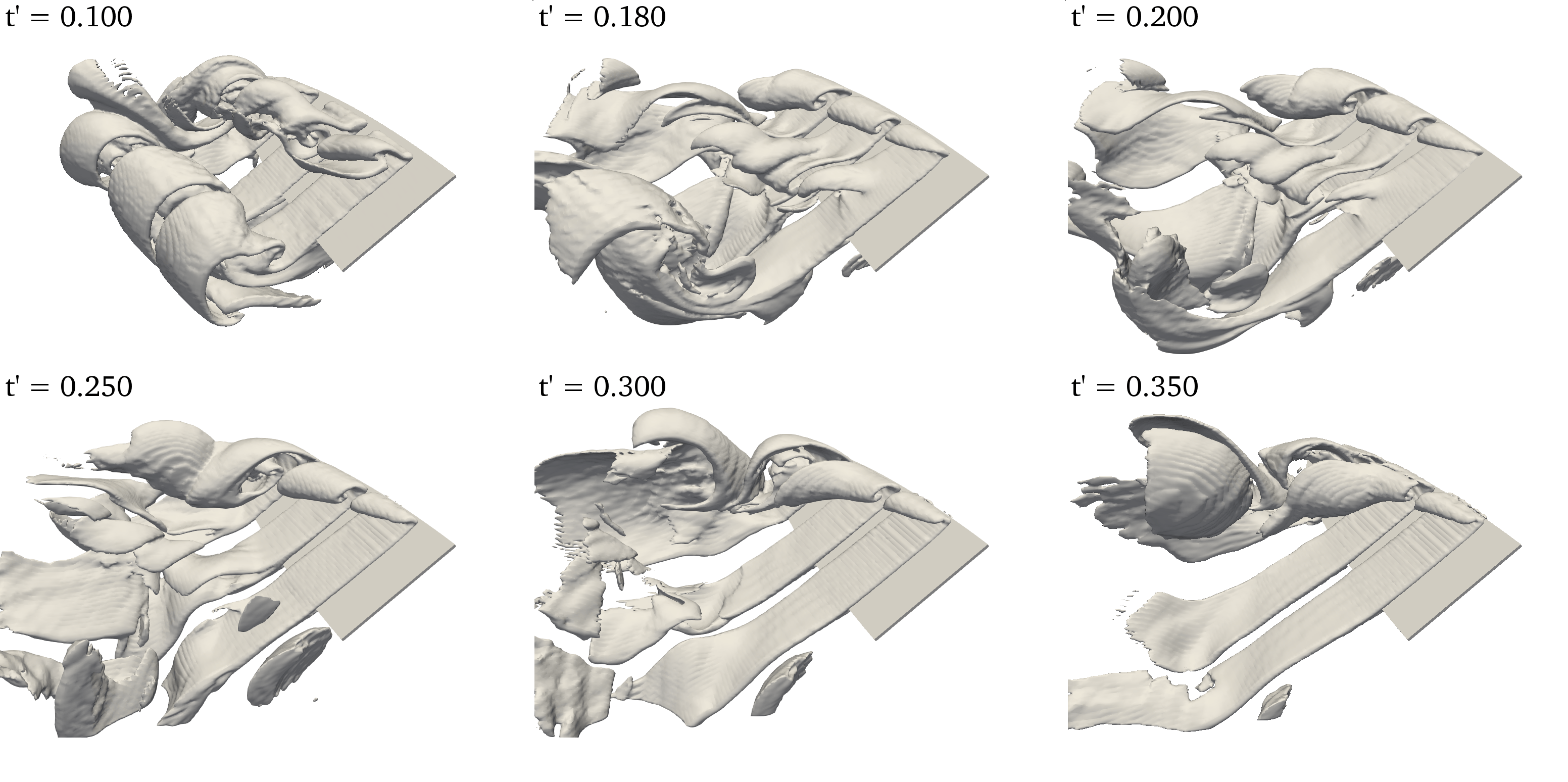}
    \end{minipage}
    \end{minipage}
\begin{minipage}{0.33\textwidth}
\begin{minipage}{0.05\textwidth} 
        \vspace{-5.1cm}
        \captionsetup{justification=raggedright,singlelinecheck=false,format=hang}
        \caption{}
        \label{img_ftleGrowth}
    \end{minipage}%
    \begin{minipage}{0.95\textwidth} 
        \centering
	 \includegraphics[width=\textwidth]{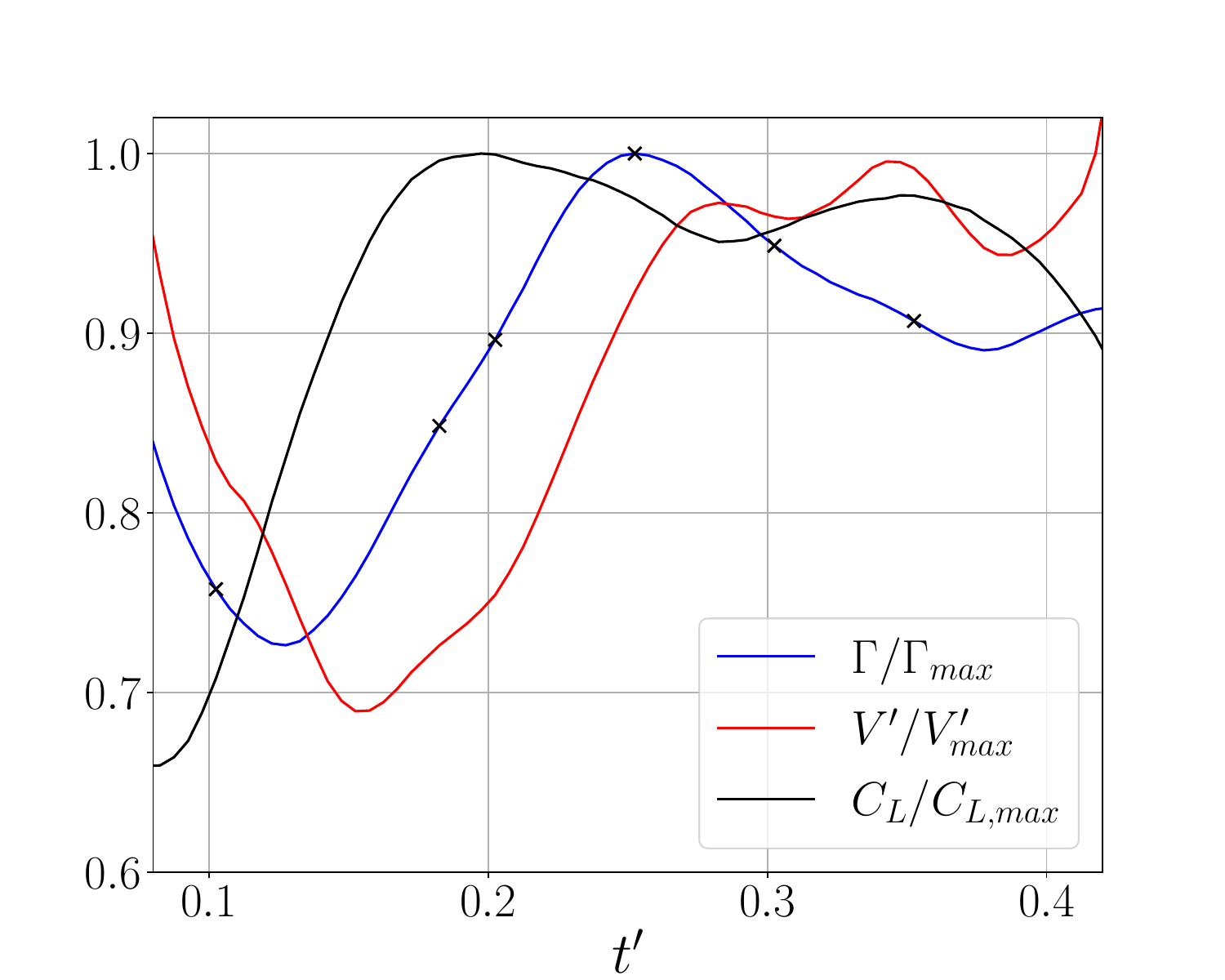}
    \end{minipage}
    \end{minipage}
\end{subfigure}
     \caption{Evolution of the LEV shape evidenced with iso-contours of the FTLE for (a) $(K_\phi,K_\alpha)=(0.99,10)$, (b) $(K_\phi,K_\alpha)=(0.5,10)$ and (c) $(K_\phi,K_\alpha)=(0.99,5)$ for $t'\in[0.1,0.35]$ evidenced with markers in Figure \ref{img_liftSide}. The time evolution of the normalized LEV circulation, LEV volume and lift is compared in (d) for $(K_\phi,K_\alpha)=(0.99,10)$.}
 \label{img_ftle}    
\end{figure*}

At the end of the acceleration phase ($t'>0.08$), the wing reaches the maximum angle of attack and a constant flapping velocity (for $(K_\phi, K_\alpha)=(0.99,10)$). During this phase, the lift increases until it saturates at $t'\sim0.3$ for the first cycle and $t'\sim0.2$ for the fifth cycle (Figure \ref{img_liftKphiVar}). Unsurprisingly, the lift is mostly generated by the suction side of the wing (Figure \ref{img_liftSide}) due to the leading-edge vortex (LEV2) which grows and then reaches a steady state. 
Figure \ref{img_ftle_K09910} shows the edges of the LEV2 using iso-contours of the FTLE fields for the 5th downstroke of the reference case from $t'=0.1$ to $t'=0.35$.
Initially, the shedding of the three starting vortices is visible and a thin shear layer connects the LEV and TEVs on the wing's upper surface.
After $t'=0.1$, the LEV2 forms and grows close to the wing. The LEV2 grows normally ($y_\alpha$), radially ($z_\alpha$) and tangentially ($x_\alpha$), absorbing most of the vorticity fed from the leading edge.  The proximal part of the LEV2 rapidly reaches its maximum size ($t'\sim0.2)$ while the distal part grows normally and tangentially until it reaches the wing's TE ($t'\sim0.3)$. 
During this time and for the whole of phase II, the LEV2 remains attached to the wing thanks to an axial flow from root to tip. The axial flow plays two roles: (1) it forms the tip vortex that was already shown to interact with the wing at the beginning of phase I and (2) it continuously removes vorticity from the LEV and thus prevents detachment due to an excessive growth \cite{Lentink2009}. This phenomenon is well known in the literature, but other stabilizing mechanisms may also occur as recently reviewed in \citet{Chen2023}. 

Figure \ref{img_ftleGrowth} illustrates quantitatively the emergence of the LEV with the time evolution of the radial circulation (along $x_\alpha$) of the LEV for the reference kinematics. Inspired from \citet{chen2020}, the circulation is computed as: 

\begin{equation}\label{eq_Gamma}
    \Gamma= - \int_{V'} \omega_{x_\alpha} dV \quad\mbox{with}\quad V': Q>0 \; \cap \; \omega_{x_\alpha}<0\,.
\end{equation}

The volume $V'$ is also clipped to keep only the LEV regions in proximity to the wing. Figure \ref{img_ftleGrowth} shows this volume in red together with the lift in black. The black markers indicate the same instants as in Figure \ref{img_ftle_K09910}.

The three curves follow a similar trend, albeit with a temporal shift.
From $t'\sim0.1$ to $t'\sim0.2$, the lift depends almost linearly on the radial circulation inside the LEV. This confirms the dominance of the LEV on the wing aerodynamics after phase I.
The circulation in turn depends linearly on the LEV volume that increases with time until $t'\sim0.3$. 

The LEV strength is known to be a function of the squared flapping velocity once a constant angle of attack is reached\cite{Dickinson1999,Eldredge2019,Krishna2018}. This is verified when the wing starts to flap smoothly ($K_\phi=0.5$) comparing the trend of the lift and $\dot{\phi}$ (Figure \ref{img_motionParametrization}).  
When the wing impulsively starts to flap ($K_\phi=0.99$), it is almost free of any attached vortices at the end of phase I but the flapping velocity is already maximal. The LEV gradually builds to reach its steady state defined by the cruise flapping velocity.
The LEV's dependency on $\dot{\phi}^2$ also explains its growth along the span according to the linear relation between the flapping velocity and the spanwise position. LEV2 has then a conical shape, from root to tip, different from the cylindrical shape of the rapidly shed LEV1 (see Figure \ref{img_wyStart}). 
At $t'\sim0.25$ and $t'\sim0.27$, the LEV circulation and volume reach a maximum slightly after the lift (Figure \ref{img_ftleGrowth}). The LEV circulation then slightly decreases, and the LEV volume oscillates around the first maximum. These trends can not be directly related to the lift oscillations that are detailed in the next section.  
Overall, the growth and saturation of the LEV are similar for different pitching dynamics (comparing Figure \ref{img_ftle_K09910} and \ref{img_ftle_K0995}).
The same conclusion is made in \citet{chen2020} who investigated the LEV formation after an impulsive start keeping a constant angle of attack. 
When reducing $K_\phi$ (Figure \ref{img_ftle_K0510}), the LEV starts growing slightly later ($t'=0.1$ and $t'=0.18$)  and reaches a larger volume and circulation which increases even after the other cases have saturated (Figure \ref{img_liftKphiVar}). This is explained by the higher cruise flapping velocity.  

\vspace{0.2cm}
\paragraph{Wing-wake interaction}
Similarly to the first phase, the influence of the leading-edge vortex on the lift is also modulated by the wake from the previous strokes (Figure \ref{img_liftKphiVar}). Three wake effects are noticed for the three kinematics in the fifth cycle compared to the first one: (1) the lift grows faster, (2) the lift saturates earlier and at a slightly lower value, and (3) the lift oscillates more. This is mostly seen on the suction side of the wing (Figure \ref{img_liftSide}), and the following section focuses on the latter for the most dynamic case. 

\begin{figure*}[!ht]\center
\begin{subfigure}{0.48\textwidth}
    \begin{minipage}{0.05\textwidth} 
        \vspace{-3.5cm}
        \captionsetup{justification=raggedright,singlelinecheck=false,format=hang}
        \caption{}
        \centering
        \label{img_ww_1}
    \end{minipage}%
    \begin{minipage}{0.95\textwidth} 
        \centering
        \includegraphics[width=0.95\textwidth]{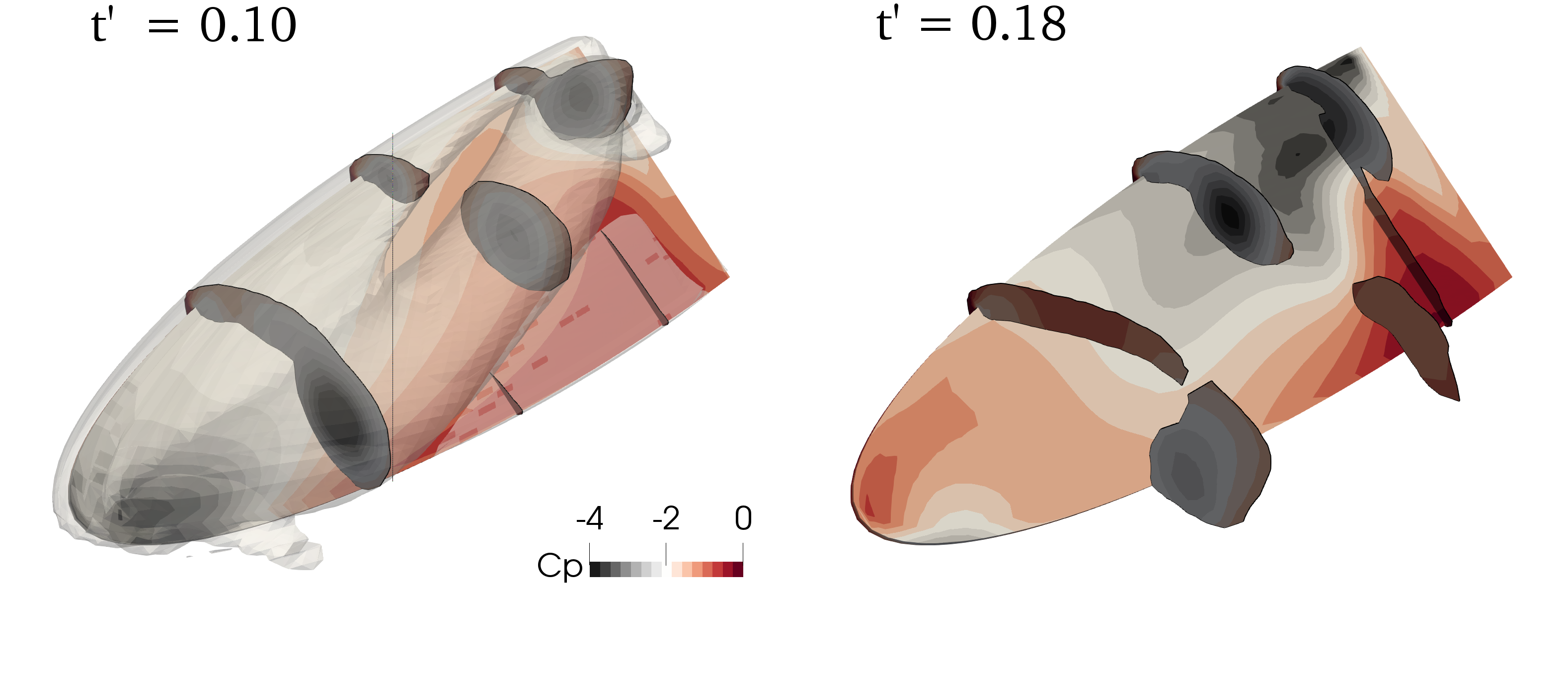}
    \end{minipage}
\end{subfigure}
\begin{subfigure}{0.48\textwidth}
    \begin{minipage}{0.05\textwidth} 
        \vspace{-3.5cm}
        \captionsetup{justification=raggedright,singlelinecheck=false,format=hang}
        \caption{}
        \centering
        \label{img_ww_2}
    \end{minipage}%
    \begin{minipage}{0.95\textwidth} 
        \centering
        \includegraphics[width=0.95\textwidth]{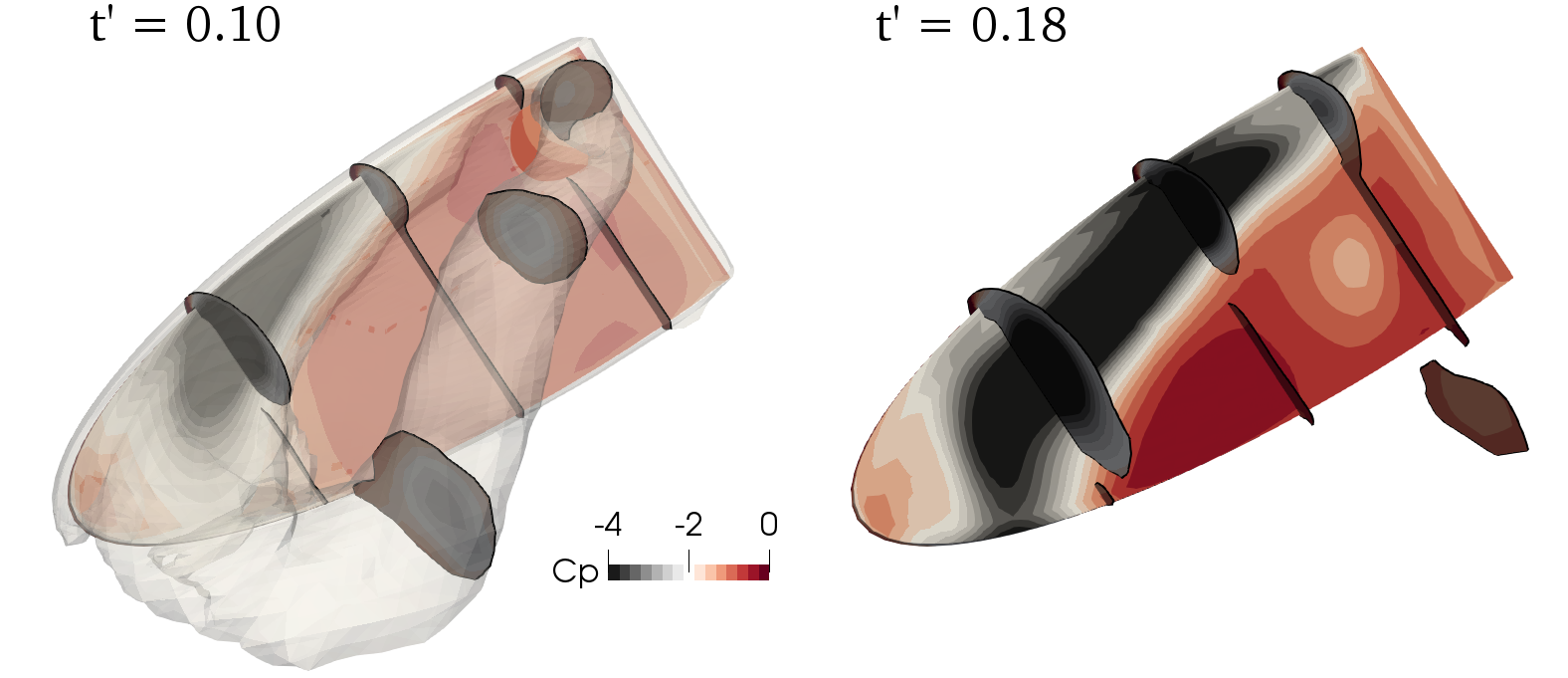}
    \end{minipage}
\end{subfigure}   
\begin{subfigure}{0.48\textwidth}
    \begin{minipage}{0.05\textwidth} 
        \vspace{-3.5cm}
        \captionsetup{justification=raggedright,singlelinecheck=false,format=hang}
        \caption{}
        \centering
        \label{img_ww_3}
    \end{minipage}%
    \begin{minipage}{0.95\textwidth} 
        \centering
        \includegraphics[width=0.95\textwidth]{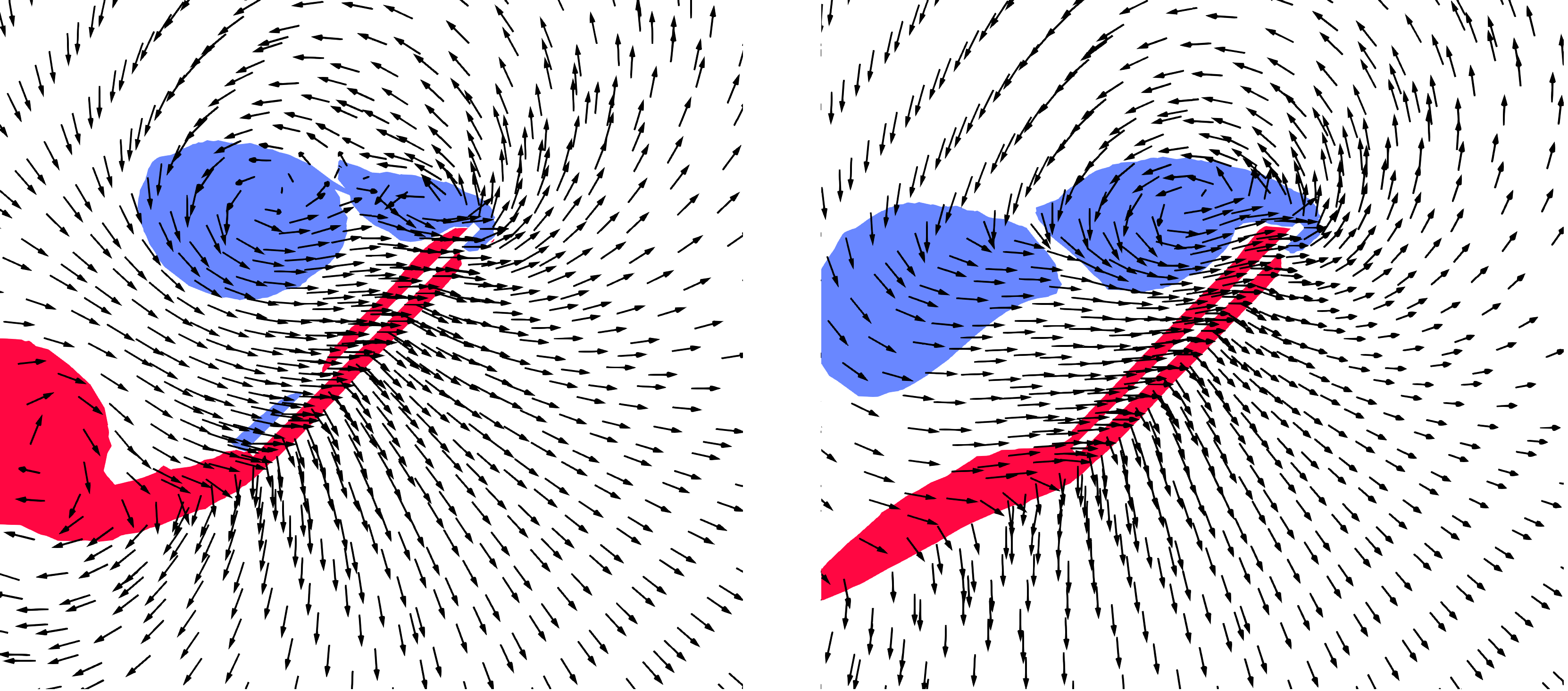}
    \end{minipage}
\end{subfigure}
\begin{subfigure}{0.48\textwidth}
    \begin{minipage}{0.05\textwidth} 
        \vspace{-3.5cm}
        \captionsetup{justification=raggedright,singlelinecheck=false,format=hang}
        \caption{}
        \centering
        \label{img_ww_4}
    \end{minipage}%
    \begin{minipage}{0.95\textwidth} 
        \centering
        \includegraphics[width=0.95\textwidth]{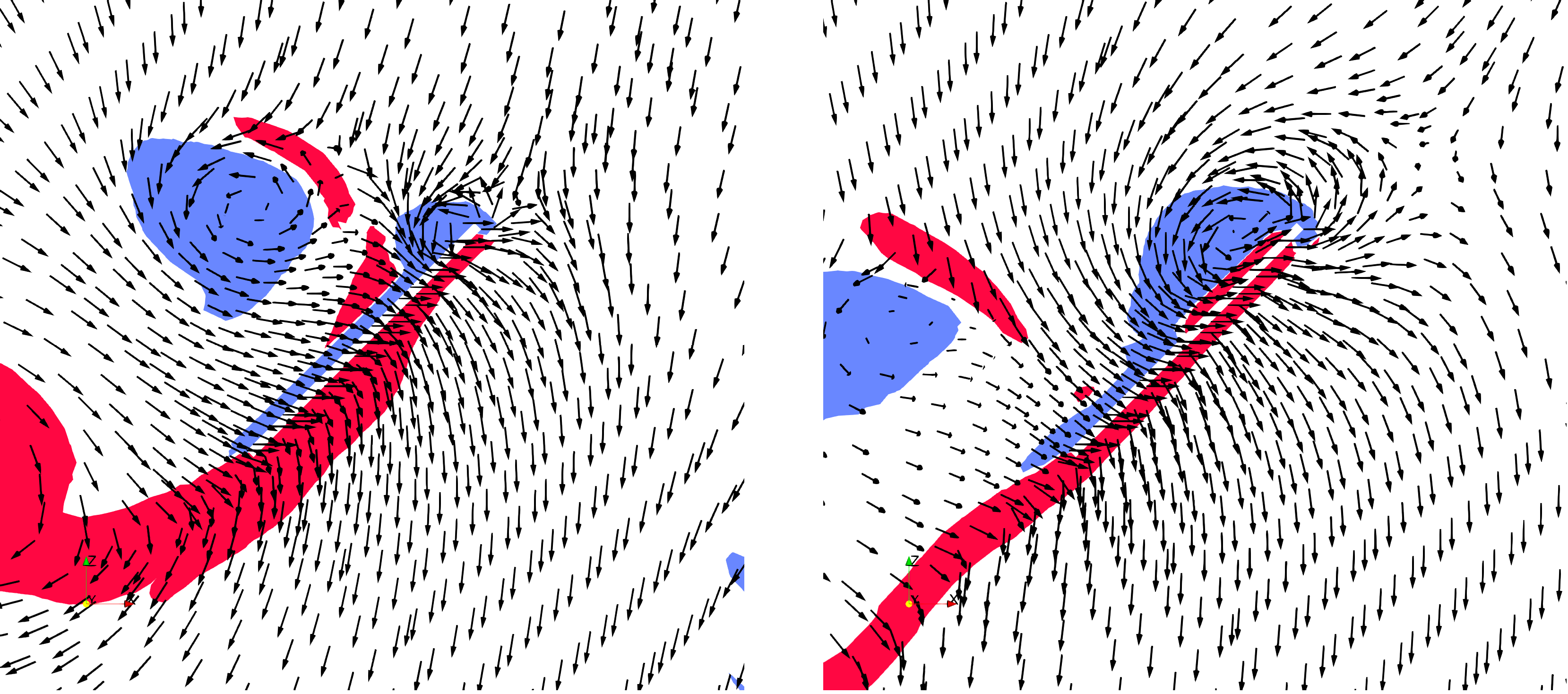}
    \end{minipage}
\end{subfigure}
     \caption{Pressure distribution on the suction side of the wing due to the LEV evidenced by $\omega_\alpha<0$ during the (a) 1st and (b) 5th downstroke at $t'=0.10$ and  at $t'=0.18$. Three slices of the LEV are also shown at $x_\alpha=[0.95R_2,R_2,1.05R_2]$. The radial vorticity and the velocity vectors are shown on the $R_2$ slice for the (c) 1st and (d) 5th cycle.}
 \label{img_wwint}    
\end{figure*}

The first wake effect is analyzed with Figure \ref{img_wwint} at instants $t'=0.10$ and $t'=0.18$ (see markers (f) and (g) on Figure \ref{img_liftSide}).
Figure \ref{img_ww_1} (first cycle) and \ref{img_ww_2} (fifth cycle) show the pressure distribution on the suction side and on three spanwise planes ($y_{\alpha} =[0.95R_2$;$R_2$;$1.05R_2]$) slicing the LEV2 highlighted by the contour of $\omega_{x_\alpha}<0$. The complete LEV2 is also shown at $t'=0.10$.
At this instant, the lift during the 1st cycle is the largest because 
the LEV2 expands on a wider tangential region at the wing tip, where it is combined with LEV1.  
The stronger coupling of the two LEVs comes from the smaller distances between them, which might not only be due to the absence of wake. It might also be due to the absence of the pitching-up rotation ($\alpha = -45^{\circ}$  to $\alpha = 0^{\circ}$) before the first cycle. 
At $t'=0.18$, LEV1 is shed further in the wake, decreasing its influence on the wing. 

\begin{figure*}[!ht]\center
\begin{subfigure}{\textwidth}
    \begin{minipage}{0.05\textwidth} 
        \vspace{-4.5cm}
        \captionsetup{justification=raggedright,singlelinecheck=false,format=hang}
        \caption{}
        \centering
        \label{img_ww_5}
    \end{minipage}%
    \begin{minipage}{0.95\textwidth} 
        \centering
        \includegraphics[width=\textwidth]{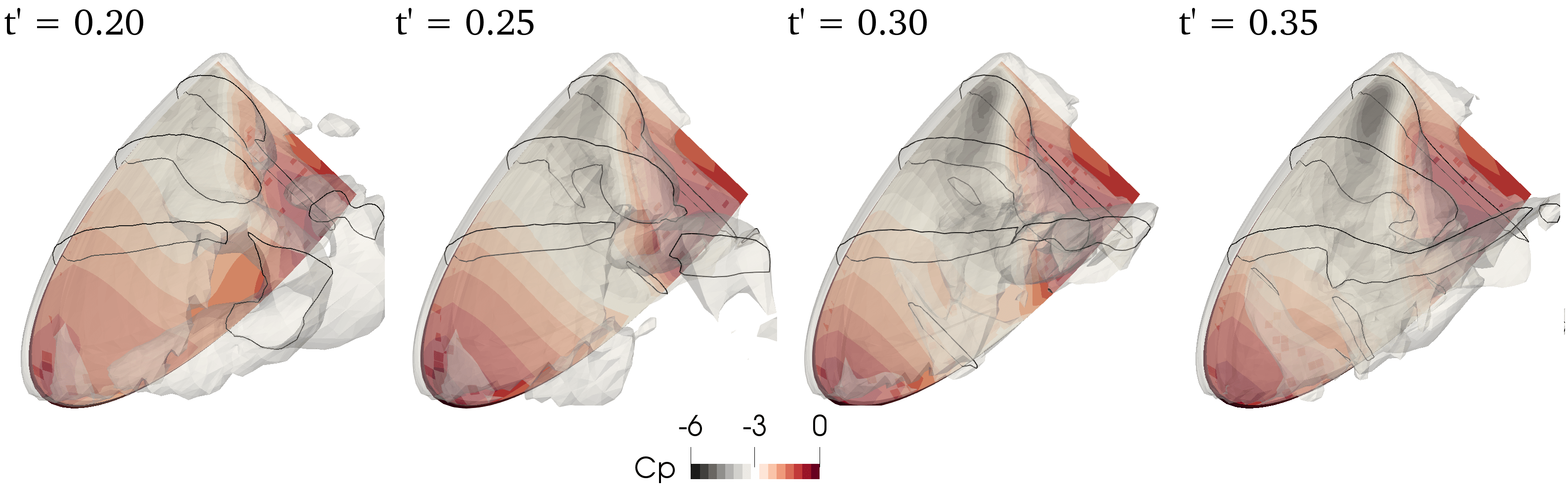}
    \end{minipage}
\end{subfigure}
\begin{subfigure}{\textwidth}
    \begin{minipage}{0.05\textwidth} 
        \vspace{-4.5cm}
        \captionsetup{justification=raggedright,singlelinecheck=false,format=hang}
        \caption{}
        \centering
        \label{img_ww_6}
    \end{minipage}%
    \begin{minipage}{0.95\textwidth} 
        \centering
        \includegraphics[width=\textwidth]{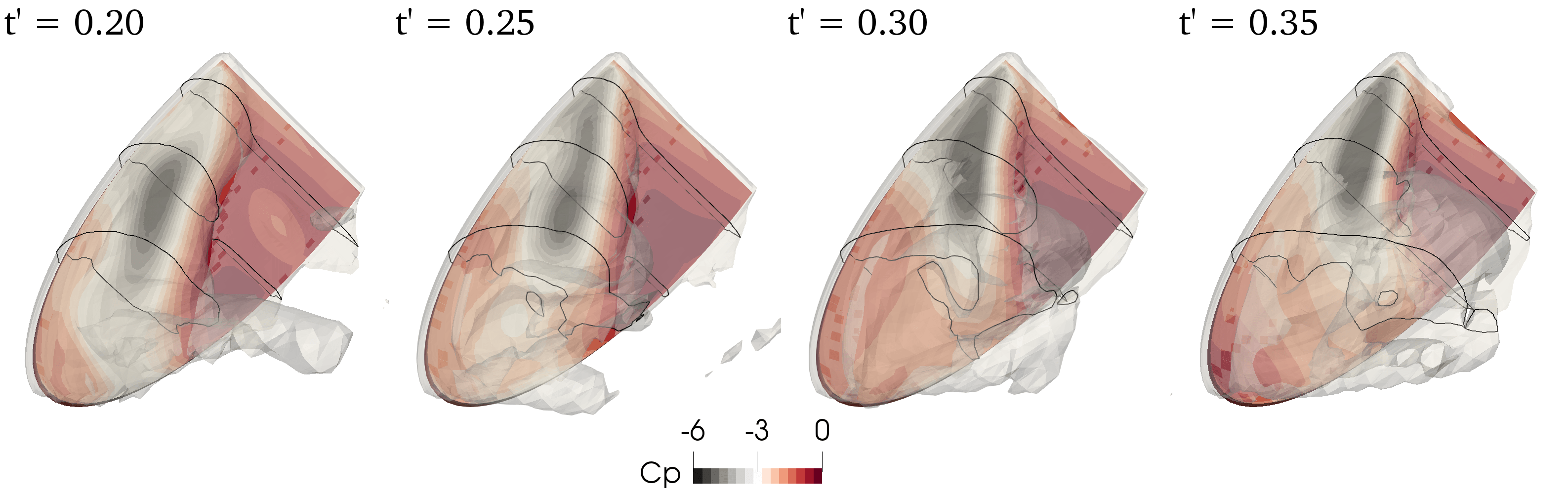}
    \end{minipage}
\end{subfigure}
     \caption{Pressure distribution on the suction side of the wing and the LEV evidenced by $\omega_{x_\alpha}<0$ during the (a) 1st and (b) 5th downstroke for $t'=[0.2,0.25,0.30,0.35]$ (see markers in Figure \ref{img_liftSide}). Three slices of the LEV are also shown at $x_\alpha=[0.95R_2,R_2,1.05R_2]$.}
 \label{img_wwint2}    
\end{figure*}

Simultaneously, the LEV2 applies more suction on the wing due to its closer proximity to the wing during the fifth cycle (Figure \ref{img_ww_2} at $t'=0.18$.).
This lift enhancement mechanism was already observed in \citet{Lee2018} in the case of a rectangular wing at $AR=3$, $Re=100$, and $2\le Ro \le 4$. The authors hypothesize that the LEV forms close to the wing due to a downwash flow, and Figure \ref{img_ww_3} and \ref{img_ww_4} provide additional details. The figures show the $R_2$ slice colored by the spanwise vorticity $\omega_{x_\alpha}$ (blue anti-clockwise and red clockwise) and with velocity vectors overlapped. 
During the first stroke (Figure \ref{img_ww_3}), the LEV1 enforces a recirculation flow between the wing and the LEV2 which may promote its lifting from the wing. 
For the 5th downstroke, the upstroke wake flow (along $y_\alpha$ and $-z_\alpha$) 
tilts the velocity vectors from the recirculation, hence reducing the effective angle of attack and constraining the LEV2 to remain closer to the wing. The wake also increases local velocities in the LEV2 and the suction pressure.\\
Building on Figure \ref{img_wwint}, Figure \ref{img_wwint2} analyzes the wake's second effect which causes the lift to saturate earlier and at a lower value at $t'\sim0.2$. The figure depicts the pressure distribution on the suction side and the LEV at three slices.
The mean pressure remains approximately constant for $t'\in[0.2,0.35]$ (Figure \ref{img_ww_6}), but the pressure distribution is nonetheless changing during these instants. 
The lowest pressure slightly increases in magnitude and moves towards the root. The decreased pressure on the distal part of the wing may be attributed to the burst of the LEV.
The downstream part of the long LEV induces a flow between the wing and the upstream part of the LEV. The recirculation makes the LEV burst, and only the upstream part remains attached to the wing (see Figure \ref{img_ww_6}).
This phenomenon is known to occur at a high Reynolds number with a small influence on the lift \cite{Harbig2013}. It also occurs during the first downstroke (Figure \ref{img_ww_5}), but later and with less influence on the lift as the LEV strength increases during most of the stroke.

\begin{figure*}[!ht]\center
\begin{subfigure}{\textwidth}
    \begin{minipage}{0.05\textwidth} 
        \vspace{-4cm}
        \captionsetup{justification=raggedright,singlelinecheck=false,format=hang}
        \caption{}
        \centering
        \label{img_oscillations_1}
    \end{minipage}%
    \begin{minipage}{0.95\textwidth} 
        \centering
        \includegraphics[width=\textwidth]{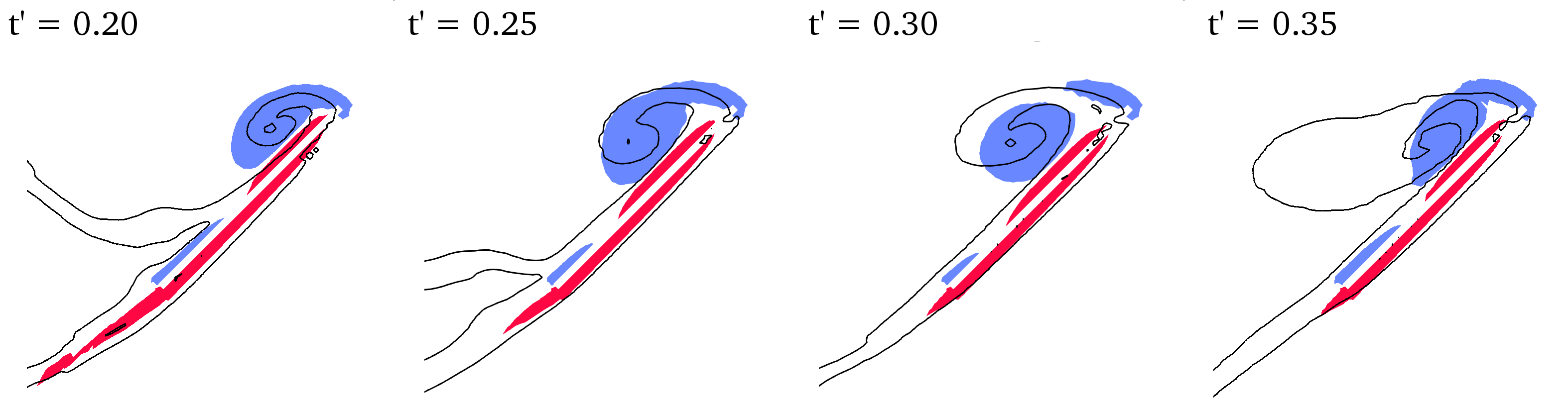}
    \end{minipage}
\end{subfigure}
\begin{subfigure}{\textwidth}
    \begin{minipage}{0.05\textwidth} 
        \vspace{-1.5cm}
        \captionsetup{justification=raggedright,singlelinecheck=false,format=hang}
        \caption{}
        \centering
        \includegraphics[width=\textwidth]{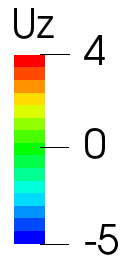}
        \label{img_oscillations_2}
    \end{minipage}%
    \begin{minipage}{0.95\textwidth} 
        \centering
        \includegraphics[width=\textwidth]{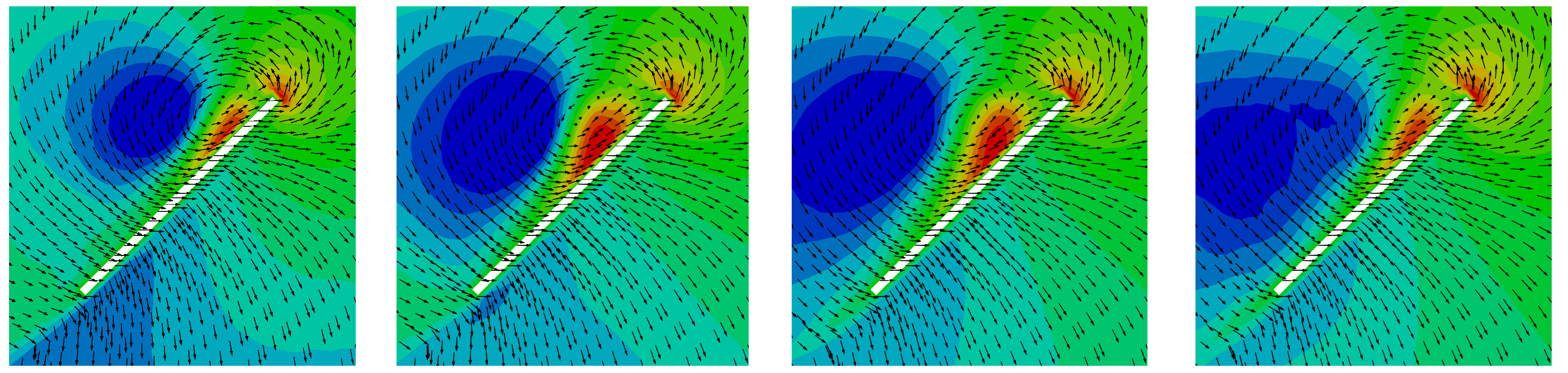}
    \end{minipage}
\end{subfigure}
     \caption{Oscillations of the LEV during the fifth downstroke highlighted by (a) the radial vorticity overlapped with an iso-contour of the FTLE and (b) the velocity $u_{z_I}$ and the velocity vectors on the $R_2$ slice.}
 \label{img_oscillations}    
\end{figure*}

The third wake effect is also attributed to the weaker reverse flow present on the proximal part of the wing. Figure \ref{img_oscillations} evidences its effect on the $R_2$ slice with (a) the radial vorticity and the iso-contours of the FTLE and, (b) the vertical velocity magnitude $u_{z_I}$ superimposed with velocity vectors.
The LEV is growing from $t'\sim0.2$ to $t'\sim0.3$ which increases the magnitude of the reverse flow along the chord, as highlighted by the red area in Figure \ref{img_oscillations_2}. This flow pushes the LEV further from the wing along $y_\alpha$ and breaks it into two LEVs at $t'=0.3$. At that instant, the lift is minimal (Figure \ref{img_liftKphiVar}). The lift then increases as the LEV reforms and approaches the wing again ($t'=0.35$).

As a final word on the wing-wake interaction, one can notice that the wake has a similar effect to the Reynolds number on the LEV described by \citet{Chen2023}: increasing the Reynolds number makes the LEV narrower and bursts more along the span. 




\subsection{Phase I and II: EPOD}
\begin{figure}[htb]
\centering
\includegraphics[width=0.45\textwidth]{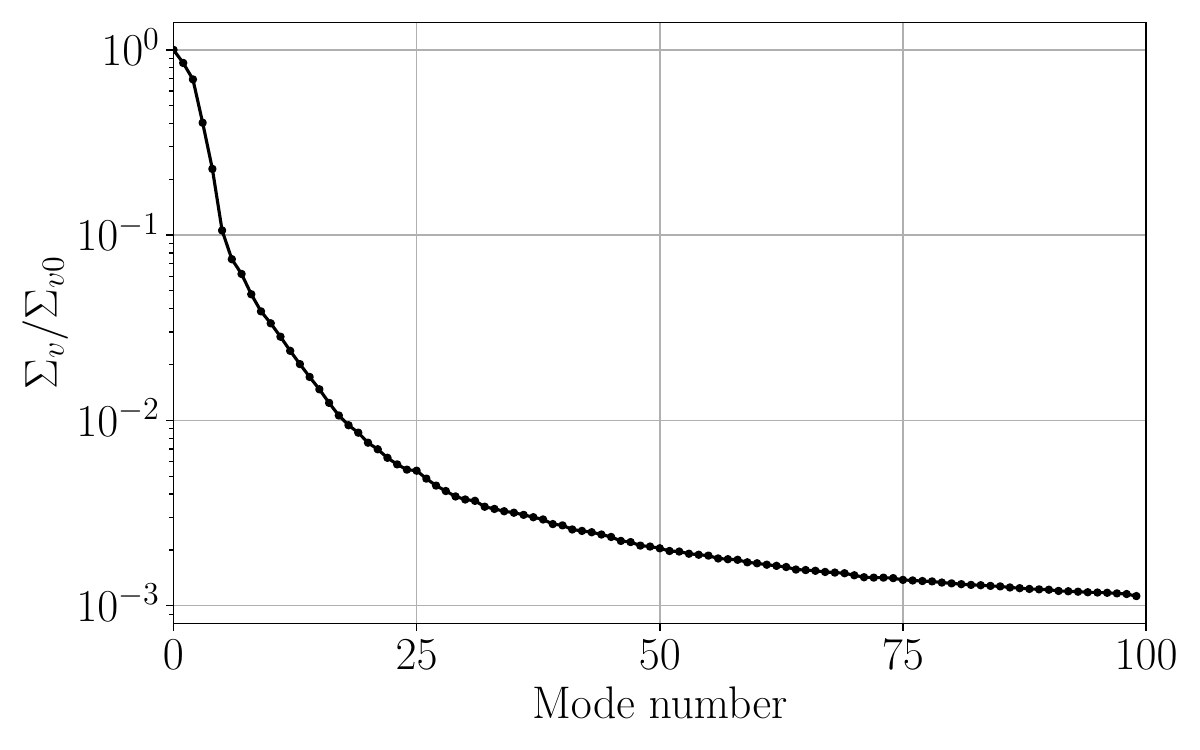}
 \caption{Normalised amplitudes of the first hundred POD modes from the velocity field.}
          \label{img_POD_Sigma}
\end{figure}

The extended POD was used to identify the link between the different flow field patterns analyzed in the previous sections and the pressure field on the wing surface.
The analysis focuses only on the most dynamic case, which generates very different aerodynamic mechanisms. 
The amplitude of the POD modes is shown in Figure \ref{img_POD_Sigma} normalized by the amplitude of the first mode. The first six modes account for $85\%$ of the total flow energy.  

\begin{figure*}
\centering
\includegraphics[width=\textwidth]{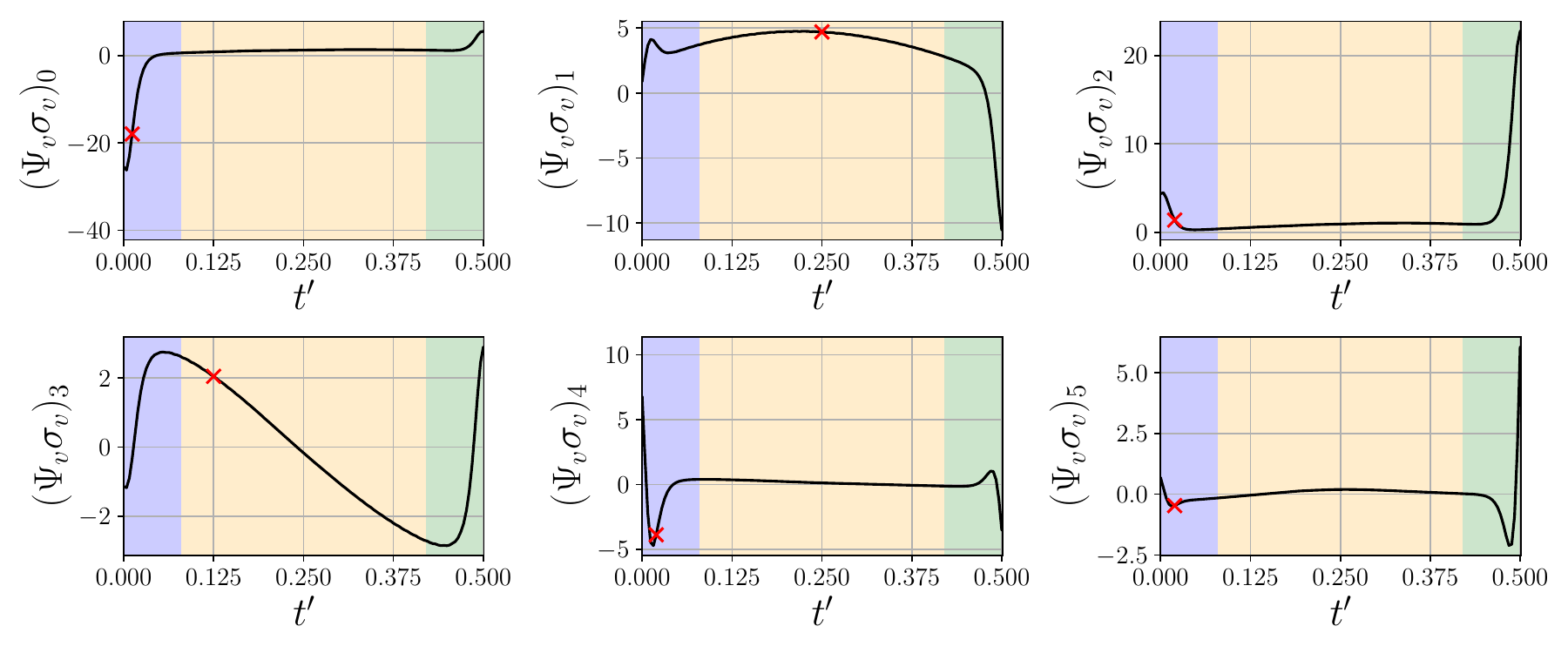}
 \caption{Temporal structures multiplied by the amplitudes for the first six modes of the POD applied on the velocity field. The markers correspond to the instants in Figure \ref{img_POD_PhiQ}.}
         \label{img_POD_Psi}
\end{figure*}

Figure \ref{img_POD_Psi} shows the products $\sigma_r \psi_r(t)$ to highlight the time evolution (weighted by the associated amplitude) of each mode as a function of time.
All modes peak during phase I and stabilize close to zero for phase II, except for modes 1 and 3. Most of the energy is thus spent impulsively starting the wing's motion. 

\begin{figure*}[!ht]\center
\begin{subfigure}{0.38\textwidth}
    \begin{minipage}{0.05\textwidth} 
        \vspace{-5.5cm}
        \captionsetup{justification=raggedright,singlelinecheck=false,format=hang}
        \caption{}
        \centering
        \label{img_POD_PhiU}
    \end{minipage}%
    \begin{minipage}{0.95\textwidth} 
        \centering
	 \includegraphics[width=\textwidth]{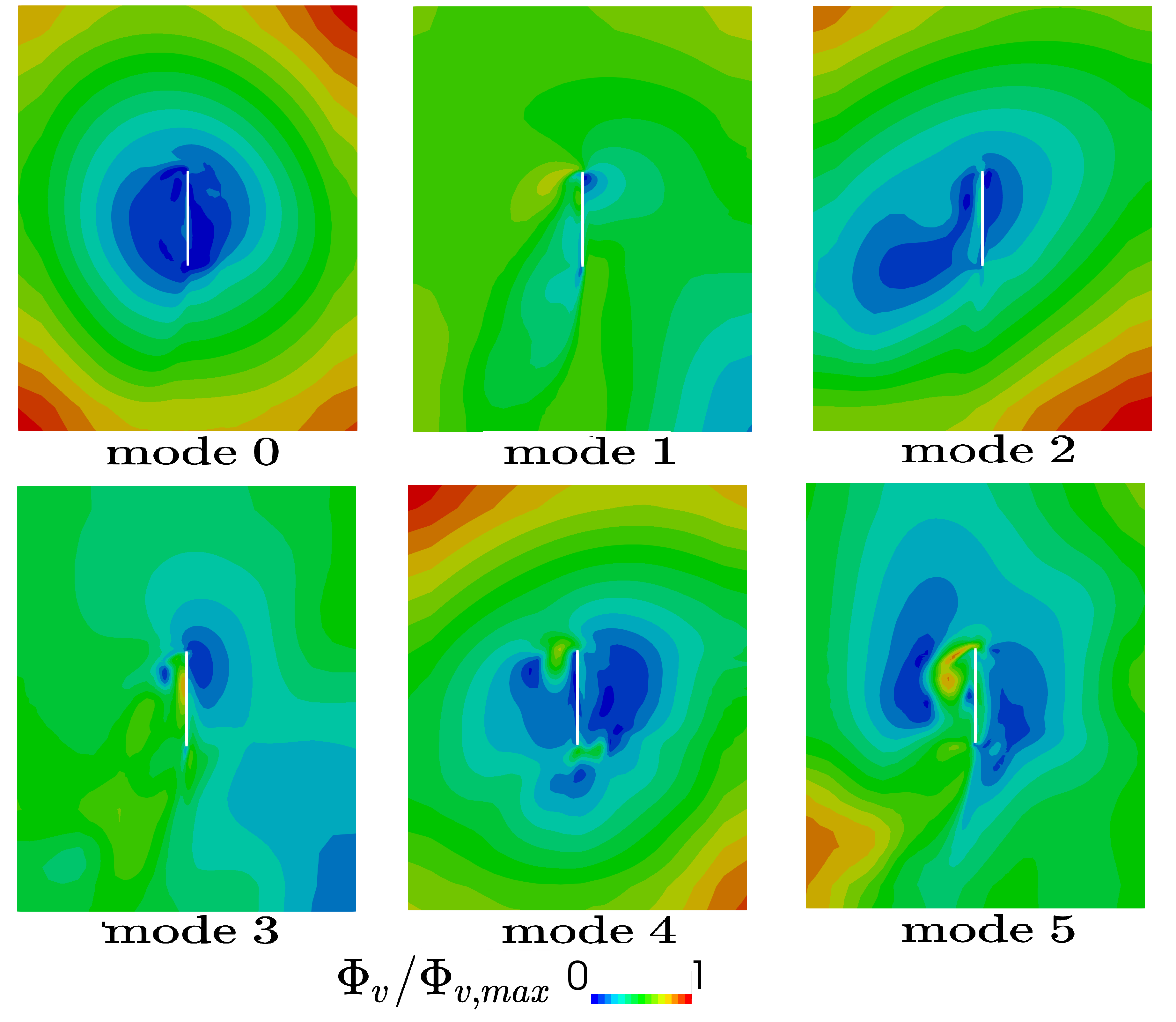}
    \end{minipage}
\end{subfigure}
\begin{subfigure}{0.58\textwidth}
    \begin{minipage}{0.05\textwidth} 
        \vspace{-5.5cm}
    \captionsetup{justification=raggedright,singlelinecheck=false,format=hang}
        \caption{}
        \centering
        \label{img_POD_PhiQ}
    \end{minipage}%
    \begin{minipage}{\textwidth} 
        \centering
	 \includegraphics[width=\textwidth]{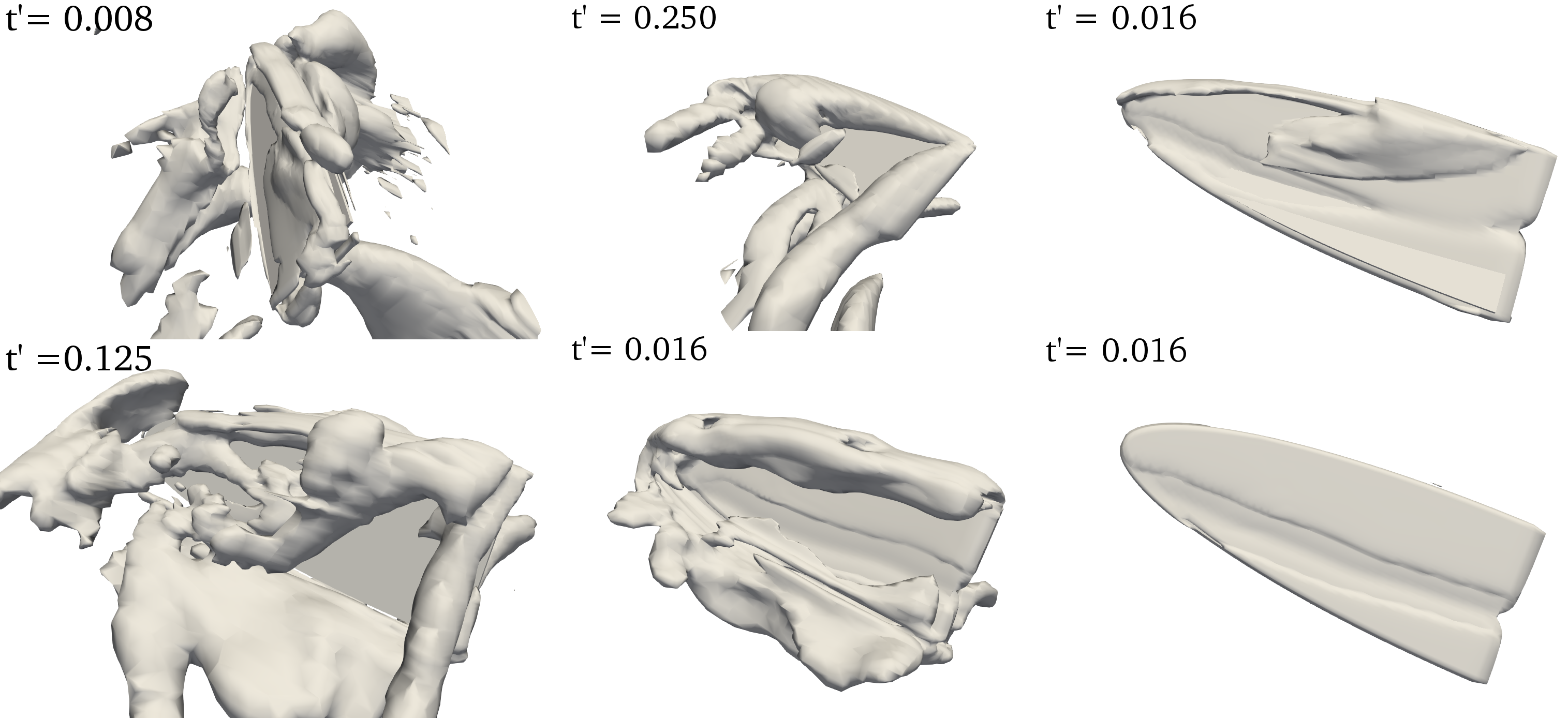}
    \end{minipage}
\end{subfigure}
     \caption{Spatial structures from the first six modes of the POD evidenced with (a) $\bm{\Phi_v}/\bm{\Phi_{v,max}}$ in the $R_2$ slice and with (b) the contours of the Q-criterion.}
 \label{img_POD_Phi}    
\end{figure*}

The largest peak (mode 0) fits well $\ddot{\phi}\cos(\alpha)$, that is the normal component of the flapping acceleration. 
The expression is classically used to model the flapping component of the added mass lift in quasi-steady models\cite{Sane2002,Lee2016} (as also used in Figure \ref{img_LiftAddM}). 
The parabolic trend of mode 1 in phase II is associated with the squared flapping velocity and hence with the formation of the LEV2. 
The role of the other modes can be further interpreted with Figure \ref{img_POD_Phi}. Figure \ref{img_POD_PhiU} shows the spatial structures on the $R_2$ plane. The wing is flapping from left to right and the spatial structures are computed on the wing frame (see section \ref{sec_FTLE}). The spatial structures are also scaled between -1 and 1.
Figure \ref{img_POD_PhiQ} shows the coherent structures at instants corresponding to the red markers from Figure \ref{img_POD_Psi}. The coherent structures are identified by the Q-criterion contours computed in the inertial frame to avoid misleading interpretations (see section \ref{sec_FTLE}).  

The leading mode (mode 0) shows a circular pattern representative of the wing pitching during phase I. The wing rotation generates the starting vortices LEV1 and TEV1. This mode also captures part of the wake from the previous stroke. 
The lowest amplitudes of the second mode (mode 1) on the suction side of the wing define the shape of the LEV2. The LEV2 has the typical conical shape and the mode also captures the root vortex.
Mode 2 stands out with a transversal flow that corresponds to the flapping kinematics. Only a few vortical structures are seen near the wing's LE.  Mode 3 adds to mode 1 in the formation of LEV2. Its temporal structure strongly differs from mode 1: it steeply increases during phase I and then decreases almost linearly during phase II, similarly to the flapping angle (Figure \ref{img_motionParametrization}). Mode 4 and 5 are weaker. They mainly capture the rotational effects at the beginning of the cycle. 

\begin{figure*}[!ht]\center
\begin{subfigure}{0.48\textwidth}
    \begin{minipage}{0.05\textwidth} 
        \vspace{-4cm}
        \captionsetup{justification=raggedright,singlelinecheck=false,format=hang}
        \caption{}
        \centering
        \label{img_POD_p_suction}
    \end{minipage}%
    \begin{minipage}{0.95\textwidth} 
        \centering
	 \includegraphics[width=\textwidth]{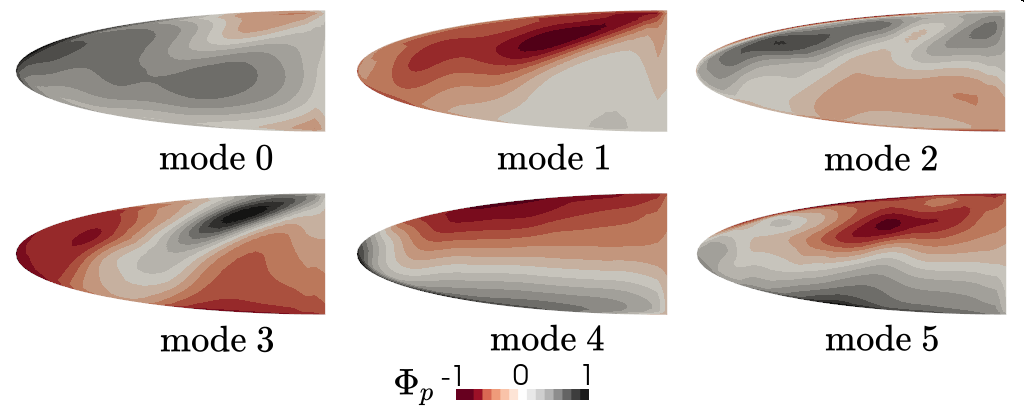}
    \end{minipage}
\end{subfigure}
\begin{subfigure}{0.48\textwidth}
    \begin{minipage}{0.05\textwidth} 
        \vspace{-4cm}
    \captionsetup{justification=raggedright,singlelinecheck=false,format=hang}
        \caption{}
        \centering
        \label{img_POD_p_pressure}
    \end{minipage}%
    \begin{minipage}{\textwidth} 
        \centering
	 \includegraphics[width=\textwidth]{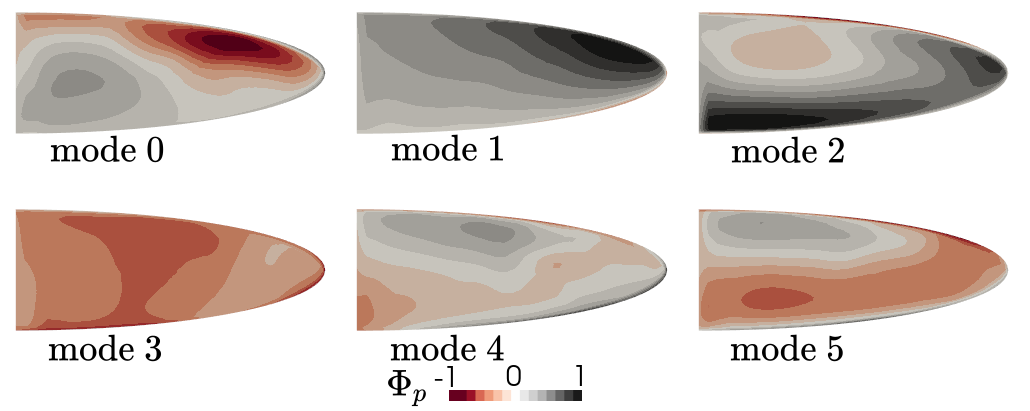}
    \end{minipage}
\end{subfigure}
     \caption{Scaled spatial structures $\bm{\Phi_p}$ of the first six modes from the pressure field shown on (a) the suction side and (b) the pressure side.}
 \label{img_POD_p}    
\end{figure*}

The pressure distribution of those six modes is shown in Figure \ref{img_POD_p_suction}  (suction side) and Figure  \ref{img_POD_p_pressure} (pressure side).  
The first and third modes are similar on the suction side as they both contain a low-pressure region included in the low-pressure ellipsoid of phase I (see section \ref{sec_phaseI}). On the pressure side, the pressure distribution from mode 0 and mode 2 resembles that of phase I during the fifth and first downstroke respectively (Figure \ref{img_wake2} and Figure \ref{img_ppressure_1st} at $t'=0.02$). The former depends on the upstroke wake characterized by the low pressure on the proximal part of the wing.
Modes 1 and 3 evidence the pressure fingerprint of the LEV2. The core of the LEV is the closest to the LE for mode 1, matching more with the shape of LEV2 during its growth. 
Modes 4 and 5 show the strongest symmetry. The wing is divided into low and high pressures along the pitching axis. 

\begin{figure*}[!ht]\center
\begin{subfigure}{0.38\textwidth}
    \begin{minipage}{0.05\textwidth} 
        \vspace{-4.5cm}
        \captionsetup{justification=raggedright,singlelinecheck=false,format=hang}
        \caption{}
        \centering
        \label{img_liftMode_0}
    \end{minipage}%
    \begin{minipage}{0.95\textwidth} 
        \centering
	 \includegraphics[width=\textwidth]{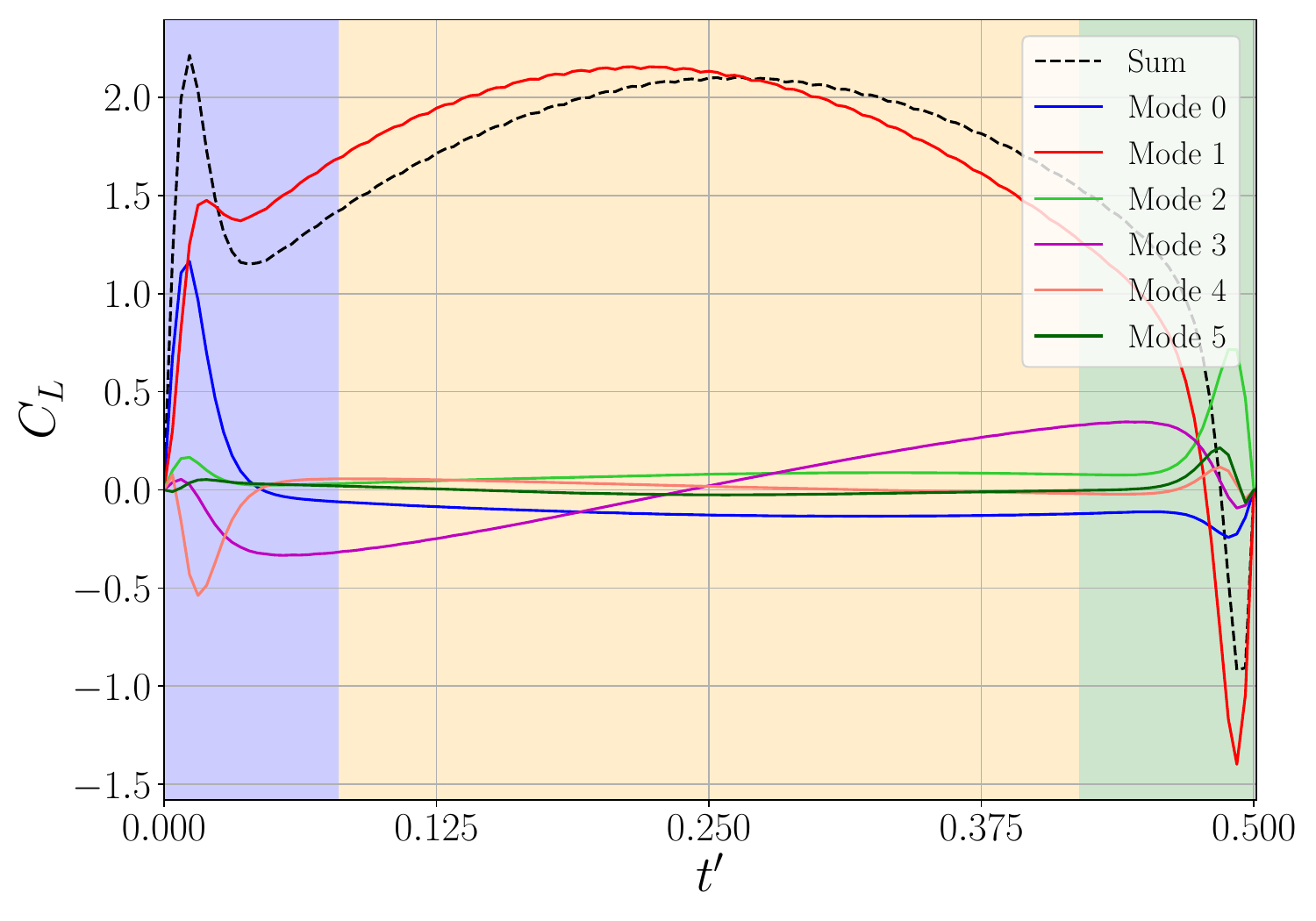}
    \end{minipage}
\end{subfigure}
\begin{subfigure}{0.58\textwidth}
    \begin{minipage}{0.05\textwidth} 
        \vspace{-4.5cm}
    \captionsetup{justification=raggedright,singlelinecheck=false,format=hang}
        \caption{}
        \centering
        \label{img_liftMode_1}
    \end{minipage}%
    \begin{minipage}{\textwidth} 
        \centering
	 \includegraphics[width=\textwidth]{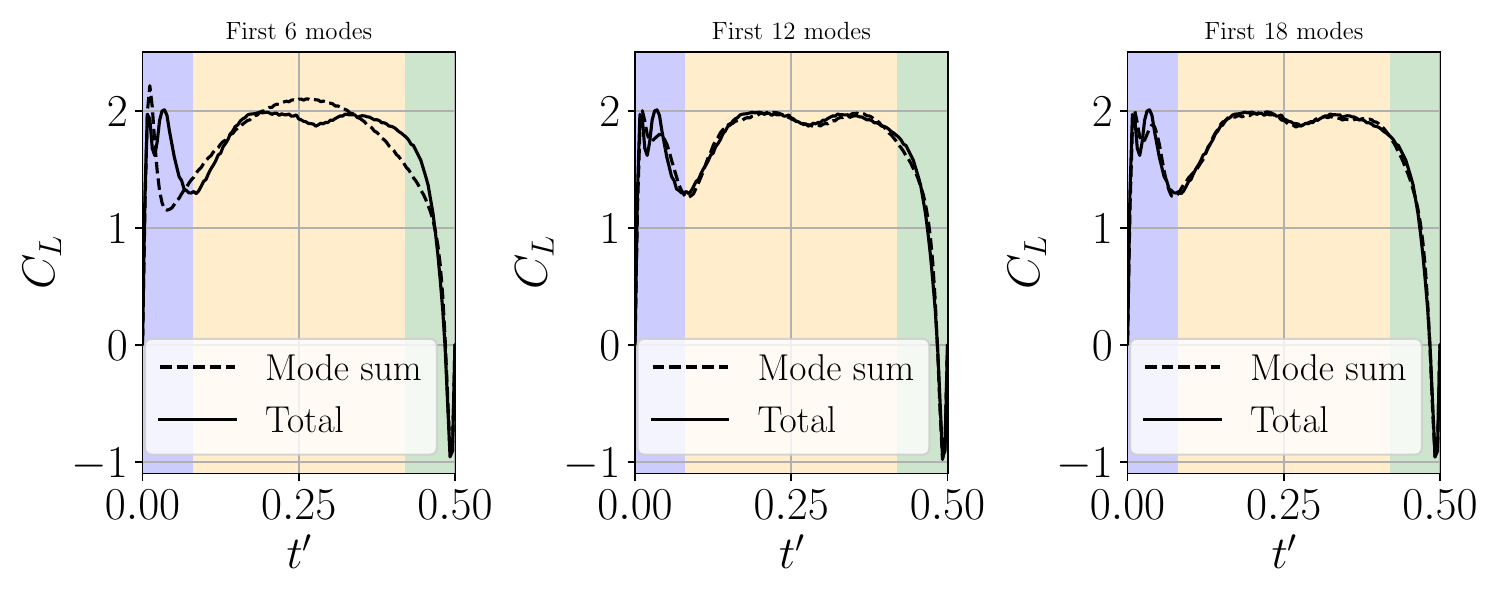}
    \end{minipage}
\end{subfigure}
     \caption{Lift as a function of time for (a) the 6 first modes and (b) the sum of the 6 first to 18 first modes.}
 \label{img_liftMode}    
\end{figure*}

Finally, the pressure distributions are integrated to estimate the lift force (equation \eqref{eq_CL}), neglecting the small forces generated by the shear stresses. The lift of the first six modes and their sum are shown in Figure \ref{img_liftMode_0}. 
By construction, the lift has the same time evolution as their temporal structures. The first and second modes dominate the lift in phase I (added mass) and the second mode dominates the lift in phase II (LEV2). Figure \ref{img_liftMode_1} shows that with only the first 6 modes the dominant trends of the lift is captured, as compared to the total lift computed with equation \eqref{eq_CL}. Adding the next 6 modes allows to capture the second lift peak and the lift oscillations during phase II. Adding additional modes results only in minor improvement of the curve fitting, hence proving the minor role of this contribution.   



\section{Conclusions}\label{sec_conclusions}


This work analyzed the unsteady aerodynamics of a semi-elliptical wing under bio-inspired flapping kinematics in the hovering regime. The flow field was simulated using high-fidelity simulations in OpenFOAM, combining an LES model and the overset method. The wing executes high-acceleration flapping and pitching at $Re\sim\mathcal{O}(10^3)$ with smoothed triangular and step waveforms. This motion generates a highly unsteady lift that results from a complex interplay of aerodynamic phenomena not fully characterized in the literature before.

This study separately analyzes the lift on the pressure and suction sides, correlating each component with surface pressure distribution and flow field structures. FTLE and Q-criterion contours were used to highlight the vortex dynamics, which were then compared to a smoother pitching motion while maintaining the flapping dynamics and vice versa. During the initial phase of the stroke, the wing dynamically flaps and pitches. The added mass mechanism was found to be the main lift contributor due to the formation of a large peak on the suction side and a smaller one on the pressure side.

Separate potential flow computations were performed in OpenFOAM to isolate the added mass effect for flapping and pitching motions. It was found that only the flapping acceleration contributes to the lift peak, which matches well with state-of-the-art quasi-steady models \cite{Lee2016}. This lift peak results from a low-pressure ellipsoid on the wing's suction side, identifiable in both potential and viscous simulations. Comparing their flow fields allowed for the identification of the roles of flapping and pitching motions in the dynamics of the three vortices generated at the leading edge, mid-section, and trailing edge. The pitching rotation of the wing also generates a small second lift peak due to a subsequent decrease and increase of pressure on the pressure side.

The wing-wake interaction modulates the lift production in three ways. The additional air resistance increases the pressure drop on the suction side; the LEV from the previous upstroke combined with the TEV increases the pressure on the distal part of the pressure side, while the downwash flow from the previous upstroke creates a pressure drop on the proximal part. The latter effect was found to be dominant, resulting in a significant second lift peak.

Following the acceleration phase, the wing maintains a constant flapping velocity and angle of attack, promoting the growth and saturation of a Leading-Edge Vortex (LEV) on the suction side, detailed using 3D contours of Finite-Time Lyapunov Exponents. The LEV grows in three directions and accumulates the vorticity generated from the LE, as evidenced by the linear dependency between the lift, the LEV circulation, and the LEV volume.

The wake from previous strokes also alters the lift in three ways when comparing the fifth and first downstroke. The LEV attaches closer to the wing, constrained by the downwash flow and the limited influence of the shed LEV; it stabilizes with a lower circulation value due to the LEV burst and increases the lift oscillations due to the varying distance between the LEV and the wing. This distance varies due to a reverse flow between the LEV and the wing.

Finally, the aerodynamics analysis was complemented by the EPOD on the velocity field and the pressure distribution on the wing's surface. The analysis confirmed that the most energetic mode is related to the added mass effect, which is the primary mechanism for lift generation during the acceleration phase of the wing. The second mode captures the LEV, which also builds on the fourth mode. The other modes describe lower amplitude structures, mostly due to the wing pitching. The projection of the wing's pressure distribution on the temporal structures allowed for the separation of the previously analyzed patterns into different modes for the suction and pressure sides, identifying their impact on the wing's load. Finally, the pressure surface integration shows that the first 12 modes are sufficient to reproduce the main dynamics in lift generation. Future work will extend the analysis to the case of flexible wings, analyzing the impact of fluid-structure interaction on the aerodynamics of flapping wings.

\section*{ACKNOWLEDGMENTS}
The first author, R. Poletti, is supported by Fonds Wetenschappelijk Onderzoek (FWO), Project No. 1SD7823N

\section*{DATA AVAILABILITY STATEMENT}
The data that support the findings of this study are available
from the corresponding author upon reasonable request.

\section{REFERENCES}
\bibliography{Poletti_et_al_2023}

\end{document}